\documentclass[aps,pre,twocolumn,showpacs,epsfig,epsf,amssymb,showkeys,tightenlines,floatfix]{revtex4-1}
\usepackage{graphicx}
\usepackage{amssymb}
\usepackage{amsmath}
\usepackage{subfigure}
\begin{document}
%\widetext

\title{ Shear induced phase separation of Wormlike micelles-nanoparticle system : formation of a long-range ordered nanoparticle structure. 
}

\author{Sk. Mubeena}
\email{mubeena@students.iiserpune.ac.in}
\affiliation{
IISER-Pune, 900 NCL Innovation Park, Dr. Homi Bhaba Road,  Pune-411008, India.\\
}
\date{\today}
\begin{abstract}
    Shearing of nanocomposites has shown to produce ordered nanostructures and orientation in the polymeric matrix. Due to the importance of the formation of an ordered structure of nanoparticles in a polymeric matrix to device wearable, flexible and photonic materials, its useful to get an insight of the shear-induced ordering and other properties of the polymeric nanocomposites. In this paper, we use a model Wormlike micellar matrix  to study the arrangement and morphology of nanoparticles in the composite when the system is sheared using a hybrid multiparticle dynamics and molecular dynamics simulation technique. We observe a shear-induced phase separation of nanoparticles as well as an increase in the long-range ordering of the nanostructures as a result of shear. We also show that in order to form a well-packed and highly ordered structure, the nanoparticles with a smaller size should be preferred.

\end{abstract}
\maketitle
\section{Introduction}

       Formation of an ordered nanostructure results in useful material properties for a wide range of systems ranging from colloids to granular media to viscoelastic systems and their composites. A low dense dispersion of nanoparticles (NPs) in a polymeric matrix enhances its mechanical properties to a great extent ~\cite{nakamura1992effect,moloney1984fracture,leblanc2002rubber} while a high density of NPs in a polymer nanocomposite with regularly packed NPs is reported to produce interesting photonic properties in the material ~\cite{braun2006introducing,van1998opals}. Moreover, there is an increasing demand for flexible and novel photonic materials that needs to study the ordered structures in dense elastomers ~\cite{gates2009flexible,kim2012large,mannsfeld2010highly}. Application of shear has shown to produce a strong ordering and orientation in such systems ~\cite{chen1994rheological}. 

     There are many reports where transient structural ordering in a colloidal suspension is produced using oscillatory shear ~\cite{ciamarra2005shear,fan2011phase,fernandez2013microscopic} like layers, strings or crystallites, etc ~\cite{besseling2012oscillatory}. A shear-induced orientation of NPs is found in polymer composites with carbon nanotubes or graphene with enhanced composite properties ~\cite{zheng2016dispersion}. A high aggregation of nanorods with a strong flow is found in a molecular dynamics simulation study by Park et. al. ~\cite{park2014controlling} while a theoretical study by Stephanou ~\cite{stephanou2015flow} shows an increased miscibility between NPs and polymer chains due to shear flow. A Brownian Dynamic study by Yamamoto et. al. ~\cite{yamamoto2012computational} on the disc-like particles in a shear flow shows that the degree of orientation of the particles is promoted by the entanglement in the polymer chains. Thus, the shearing force provides a generic tool to produce order in such viscoelastic media with a promising way for manufacturing novel photonic or flexible materials.

    The system of Wormlike micelles (WLM) form an interesting viscoelastic polymeric matrix that exhibits fascinating properties like shear-thinning, shear-thickening and shear-banding, etc ~\cite{hoffmann1981viskoelastische,rehage1982shear,forster2005shear,rehage1991viscoelastic}. Unlike polymers, Wormlike micelles (WLM) can break and join at room temperature producing an exponential distribution of their chain lengths ~\cite{turner1990relaxation}. Similar to polymeric matrices, the addition of NPs show thickening ability on WLMs and enhanced rheological properties ~\cite{zhao2017can}. An increase in viscosity is observed by Shashkina et. al. ~\cite{shashkina2005rheology} on the addition of NPs to WLMs. When silica NPs are added to WLMs formed by CTAB and sodium nitrate, an increase in the zero-shear viscosity and the relaxation time is observed by Nettesheim et. al. ~\cite{nettesheim2008influence}. A further study conducted on the same system by Hegelson using cryogenic transmission electron microscopy (cryo-TEM) indicates the formation of the micelle-nanoparticle junction as a crosslink between micelles ~\cite{helgeson2010formation}. There exist many more studies that show the enhanced rheological properties of WLMs on the addition of NPs ~\cite{fan2015nanoparticles,pletneva2015viscoelasticity,fei2017experimental}. However, the literature lacks the information that enlightens us the microstructural organization of the system that leads to the enhancement in the system properties.

    In this paper, we study the effect of shear on the WLM-NP composite using molecular dynamics simulation technique. We show the formation of a long-range ordered structure of nanoparticles as a result of shear. We also report the shear induced phase separation of nanoparticles and compare the study for different shear rates and size of nanoparticles.

\section{Model and method}

\subsection{Modelling Wormlike micelles}

    We use the same model of Wormlike micelles that is used in our earlier studies in \cite{mubeena2015hierarchical,arxive1,arxive2,arxive3}. We repeat the potentials used in this model for the convenience of the readers. 

(1) $V_2$: Two body attractive L-J potential modified by an exponential term.\\
\begin{equation}      
        V_2 = \epsilon [ (\frac{\sigma}{r_2})^{12} - (\frac{\sigma}{r_2})^6 + \epsilon_1 e^{-a r_2/\sigma}]; 
\, \forall  r_2 < r_c.
\label{eq1}
\end{equation}

where, $r_2$ is the distance between monomers. The exponential term creates a potential barrier at $r_2=1.75\sigma$ for breaking (joining) the monomers from (to) a chain. We keep $\epsilon=110k_BT$, the cutoff distance $r_c=2.5\sigma$, $\epsilon_1=1.34\epsilon$ and $a= 1.72$. \\

(2) $V_3$: Three body potential to model semiflexibility of chains.\\
 For any 3 monomers with a central monomer bonded with two other monomers at a distance $r_2$ and $r_3$ and forming an angle $theta$ at the central monomer, the following 3body potential adds the semiflexibility to the chains,

\begin{equation}
V_3 = \epsilon_3 (1 - \frac{r_{2}}{\sigma_3})^2(1 - \frac{r_{3}}{\sigma_3})^2 \sin^2(\theta); 
\, \forall r_{2},r_{3} < \sigma_3. 
\label{eq2}
\end{equation}

where, $\epsilon_{3}=6075k_{B}T$, and the cutoff distance $\sigma_3=1.5\sigma$. \\

(3) $V_4$: Four body potential to avoid branching.\\

    To avoid branching, a four body potential is used to repel any chain trying to form a branch,

\begin{equation}
V_4 = \epsilon_4 (1 - \frac{r_{2}}{\sigma_3})^2(1 - \frac{r_{3}}{\sigma_3})^2 \times V_{LJ}(\sigma_4,r_4) 
\label{eq3}
\end{equation}

where, $r_2$ and $r_3$ are the distances of the two bonded monomers from the central monomer in a chain and $r_4$ is the distance from the monomer attached to the other chain that needs to be repelled. This is a shifted L-J potential with only the positive part. The cutoff distance for this potential $\sigma_4$ is chosen such that $ \sigma_3 < \sigma_4 < r_c$ and is fixed at $\sigma_4=1.75\sigma$. This model has been already used to show the Iso-Nem transition and the exponential length distribution of the polymeric chains confirming the characteristic properties of the Wormlike micellar/equilibrium polymeric system. Please refer ~\cite{mubeena2015hierarchical,arxive1} for a detailed description of the model and its successful implementation.

\subsection{Modelling nanoparticles} 

The nanoparticles are modelled as attractive L-J particles with the ineracting potential given by,

\begin{equation}
V_{2n} = \epsilon_n[(\frac{\sigma_{n}}{r_n})^{12} - (\frac{\sigma_{n}}{r_n})^6],      \forall     r_n   <=   r_{cn}
\end{equation}

            The cutoff distance $r_{cn}$ is set at $r_{cn}=2\sigma_n$. These nanoparticles interact with monomers via a repelling potential $V_{4n}$ which is a shifted Lennard-Jones potential given by,

\begin{equation}
V_{4n} = \epsilon_{4n}[(\frac{\sigma_{4n}}{r_{mn}})^{12} - (\frac{\sigma_{4n}}{r_{mn}})^6],  \forall r_{mn} <= 2^{1/6}\sigma_{4n}
\end{equation}

Where, $r_{mn}$ indicates the distance between monomers and nanoparticles with the parameter $\sigma_{4n}$ indicating the centre-to-centre distance between the particles. The value of $\sigma_{4n}$ is always kept at $\sigma/2+\sigma_n/2$ which is the minimum possible value for a given value of $\sigma_n$ ~\cite{arxive2,arxive3}. The value of the strength of the repulsive interaction is fixed at $\epsilon_{4n}=30k_BT$.

\section{Method}

    We use a hybrid method where the fluid is simulated using the Multiparticle collision dynamics (MPCD) technique the equilibrium polymers and nanoparticles which are evolved using Molecular dynamic method (MD) interact with the fluid particles by coupling the MPCD with the MD method. To simulate fluid, the simulation box is divided into unit cells with side length a(=1). The fluid particles are sorted into these cells and the particle density is fixed as 10 particles per unit cell. The MPCD method to simulate the effect of hydrodynamics consists of two steps:

1) Streaming step: In this step, the position of a particle i at any time t, with a velocity $\bar{v_i(t)}$ is updated to a time t+h according to,

\begin{equation}
\bar{r_i}(t+h) = \bar{r_i}(t) + h\bar{v_i}(t);
\,  \forall i=1,...,N_s.
\end{equation}

\begin{figure*}
\centering
\includegraphics[scale=0.2]{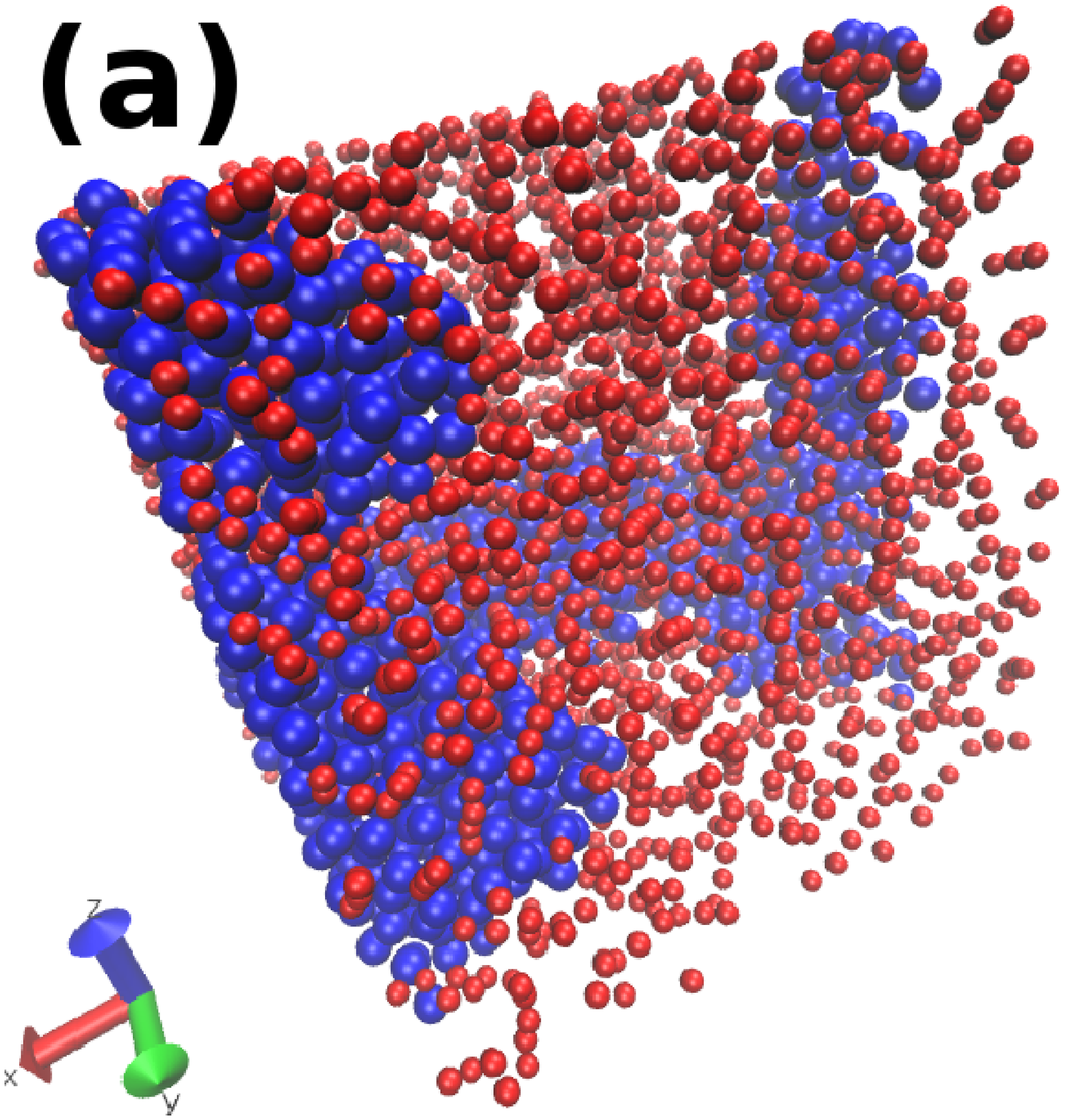}
\hspace{0.9cm}
\includegraphics[scale=0.2]{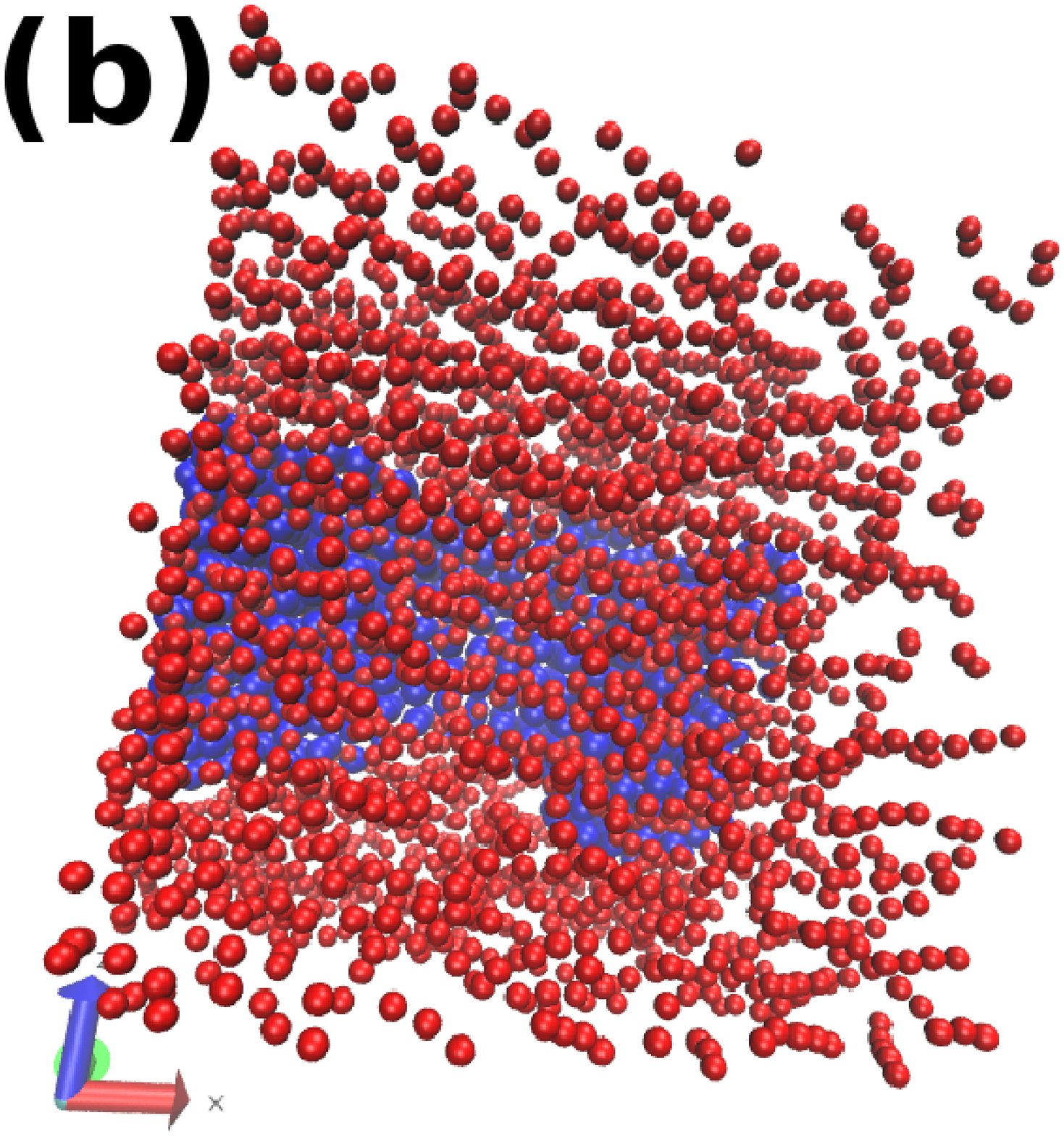}
\hspace{0.9cm}
\includegraphics[scale=0.2]{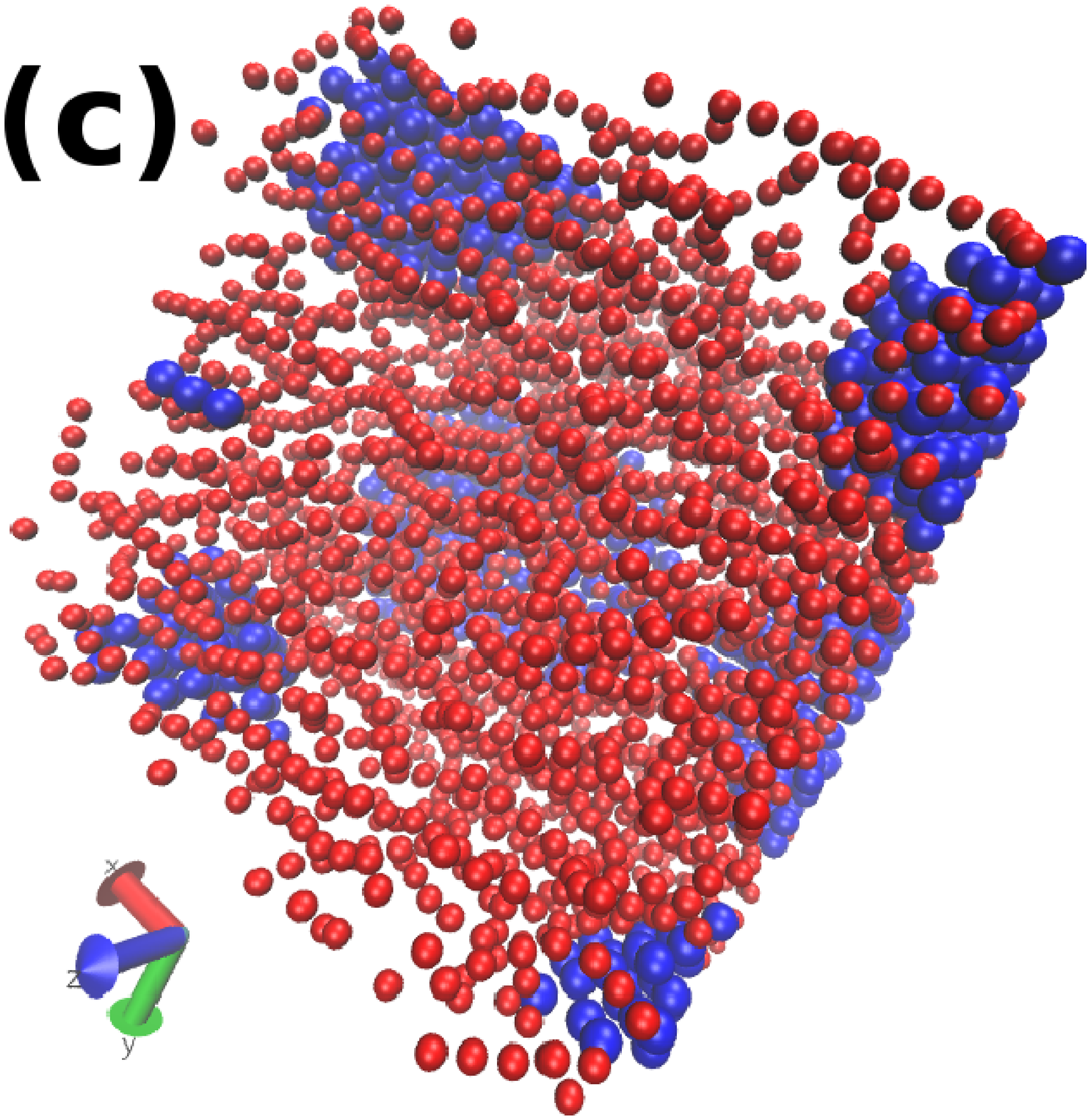}\\
\includegraphics[scale=0.2]{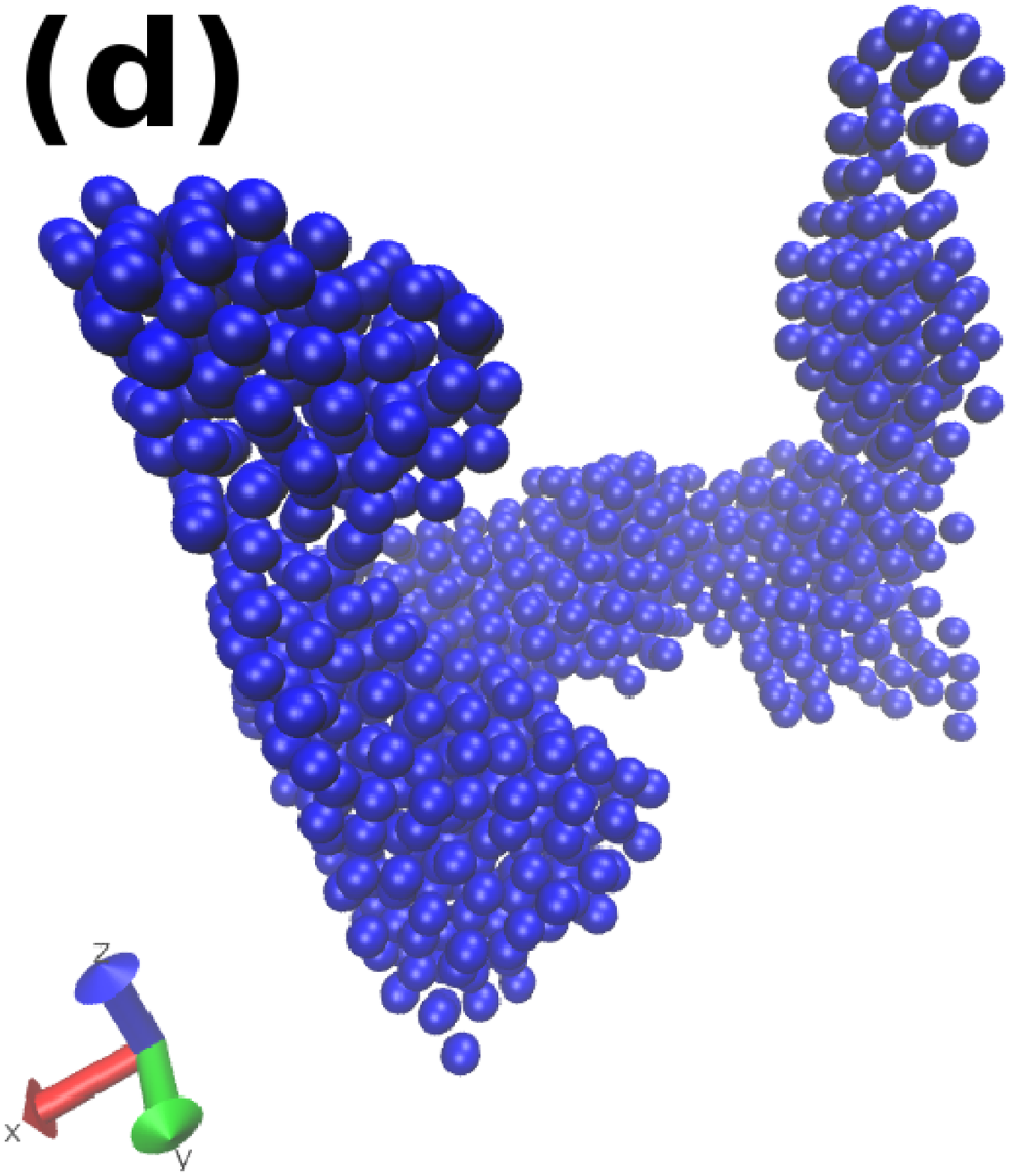}
\hspace{0.9cm}
\includegraphics[scale=0.2]{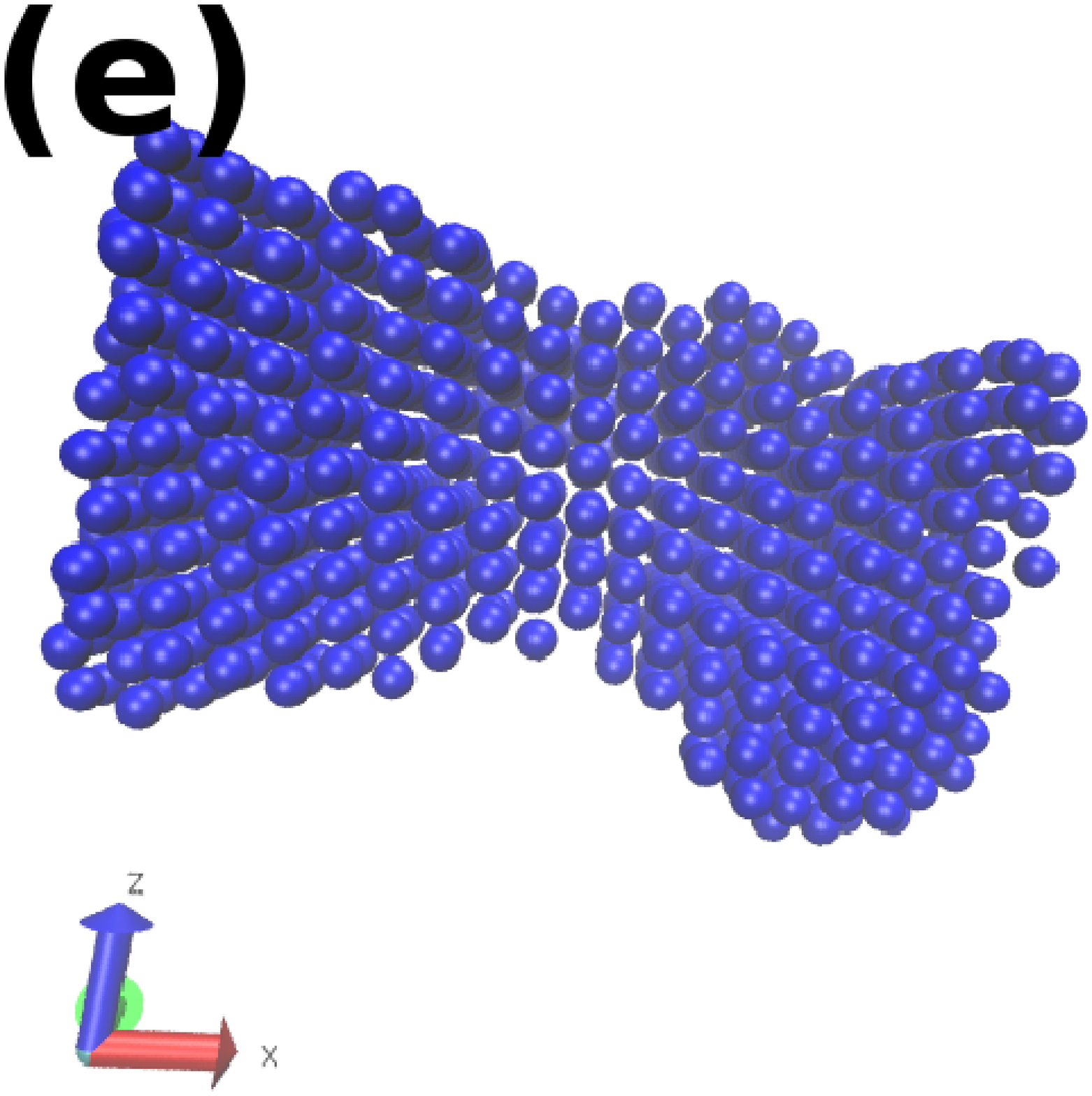}
\hspace{0.9cm}
\includegraphics[scale=0.2]{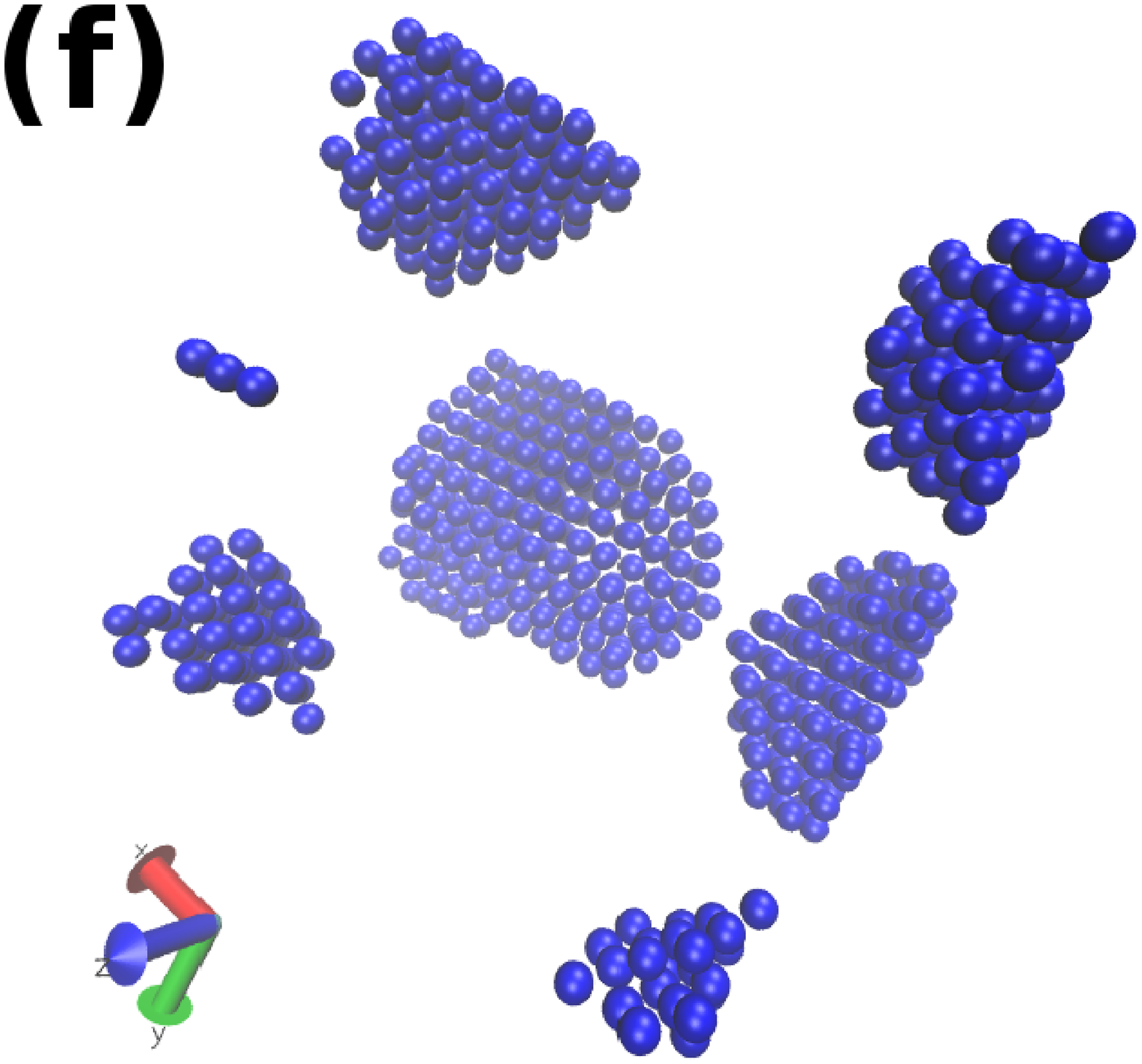}
\caption{ The figure shows the snapshots of Wormlike micelles-nanoparticle system (top row) and only the nanoparticles (bottom row) for different values of shear rates (a) $0.001\tau^{-1}$, (b) $0.01\tau^{-1}$ and (c) $0.03\tau^{-1}$. All the snapshots are for same size of nanoparticles $\sigma_n=1.65\sigma$ and $\rho_m=0.089\sigma^{-3}$. The figure shows that with increasing shear rate the NP structure transform from percolating to phase separated structure and finally to non-percolating clusters.}
\label{sh_rate}
\end{figure*}

2) Collision step:   In the collision step, the centre of mass velocity is calculated in each cell and the relative velocity (w.r.t centre of mass velocity) of the particles present in that cell are rotated around a randomly chosen axis by an angle $\alpha=130^\circ$. 

\begin{equation}
        \bar{v_i}(t+h) = \bar{v_i}(t) + (\bar{R}(\alpha)-\bar{E})(\bar{v_i}(t)-\bar{v}_{cm}(t))
\end{equation}

where, $\bar{R}(\alpha)$, $\bar{E}$ and $\bar{v}_{cm}$ are the rotation matrix, the unit matrix and the centre of mass velocity of the cell in which the particle is present. Here, $\alpha=130^\circ$.
The rotation of velocities simulates the collision of particles with each other as an effect of hydrodynamics interactions. Therefore, by including the monomers and nanoparticles in the colision step it imparts the hydrodynamic effects on the Wormlike micelle-nanoparticle system. 

Therefore, in a collision cell c with ${N_c}^m$ monomers of mass $m_m$ and ${N_c}^s$ solvent particles of mass $m_s$, the centre of mass velocity is calculated to be,

\begin{equation}
v_{cm}(t) = \frac{{\sum_{i=1}}^{{N_c}^s} m_s\bar{v_i}(t) + {\sum_{j=1}}^{{N_c}^m}m_m\bar{v_j}(t)}{{m_s{N_c}^s}+{m_m{N_c}^m}}
\end{equation}

The Galilean invariance is ensured by performing a random shift at every collision step. This method conserves the mass and linear momentum at every step.
 In order to impose shear flow along the x-z planes, the periodic boundary conditions along these planes are replaced by Lees-Edwards boundary conditions. A no slip boundary condition is maintained by imposing the bounce back rule which ensures that any particle colliding with the walls will be reverted in its velocity. A local Maxwellian thermostat maintains the fluid temperatur. We fix $k_BT=1$, collision time $h=0.1$, $m_s=1$ and collision cell size $a=1$. Then, the shear viscosity of the fluid yields the value around 8.7.
The unit of length is chosen as $\sigma=1=a$ and the simulation box is of $30\times 30\times 30\sigma^3$. The mass of the solvent particles kept at $m_s=1$.
For each collision step h, the molecular dynamics to evolve polymers is called for $h/h_p$ steps with $h_p=0.002$.
The length and time are scaled according to $\hat{r}=r/a$ and $\hat{t}=t\sqrt{k_BT/m_sa^2}$.

\section{Results ::}

%\begin{figure}
%\centering
%\includegraphics[scale=0.3]{vel_profile.eps}
%\caption{ The figure shows the velocity profile of the MPCD fluid generated by applying a wall and implementing bounce-back condition for the fluid particles colliding with the walls in a box of $10\times 10\times 10 \sigma^3$ with the no. density of fluid $10\sigma^{-3}$. The velocity profile generated is a Poiseuillie flow profile and confirms the method used.}
%\label{energies}
%\end{figure}

%\begin{figure}
%\centering
%\includegraphics[scale=0.3]{vel_prof_new.png}
%\caption{ The figure shows the velocity profile generated by applying Lees edwards boundary condition in the shearing direction for a shear rate of $0.02$ in a box of size $10\times 10\times 10\sigma^3$ with MPCD fluid particles density of $10\sigma^{-3}$}
%\label{energies}
%\end{figure}

\begin{figure}
\centering
\includegraphics[scale=0.3]{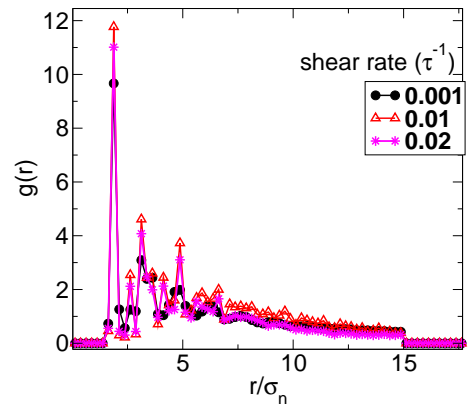}
\caption{The figure shows the pair correlation function for the NPs shown in the three snapshots in the lower row of Fig.\ref{sh_rate} differing in the values of shear rates. All the three functions show sharp peaks indicating the existence of a long-range order. However, the peaks for shear rate$=0.01\tau^{-1}$ is highest compared to other peaks indicating that the phase separated NP structure in Fig. (1(e)) has higher ordered structure. Moreover, the appearance of a peak around $2.6$ and $3.55$ indicates the hexagonal packing of the structures.}
\label{gofr_sh}
\end{figure}

\begin{figure*}
\centering
\includegraphics[scale=0.3]{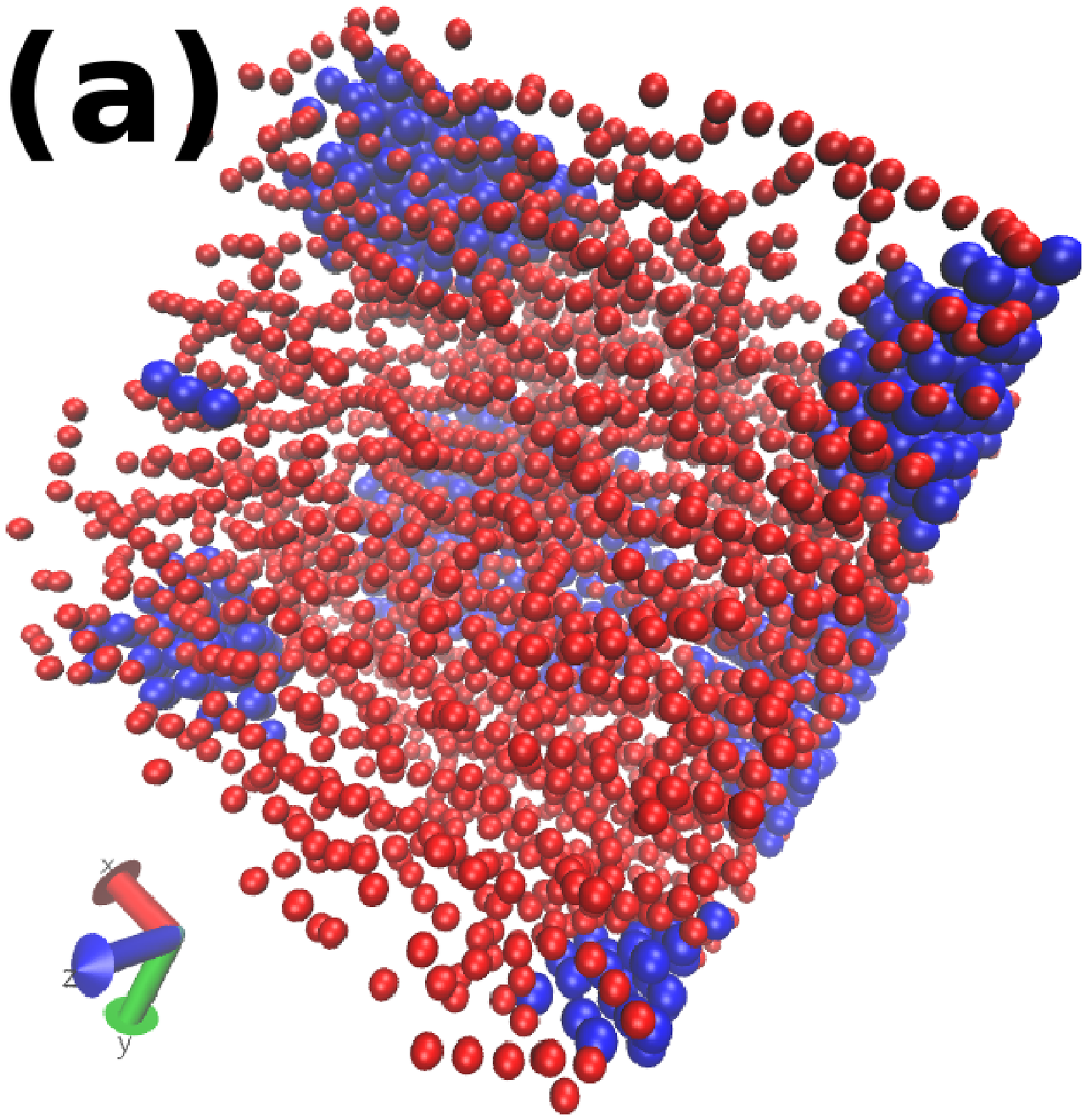}
\includegraphics[scale=0.3]{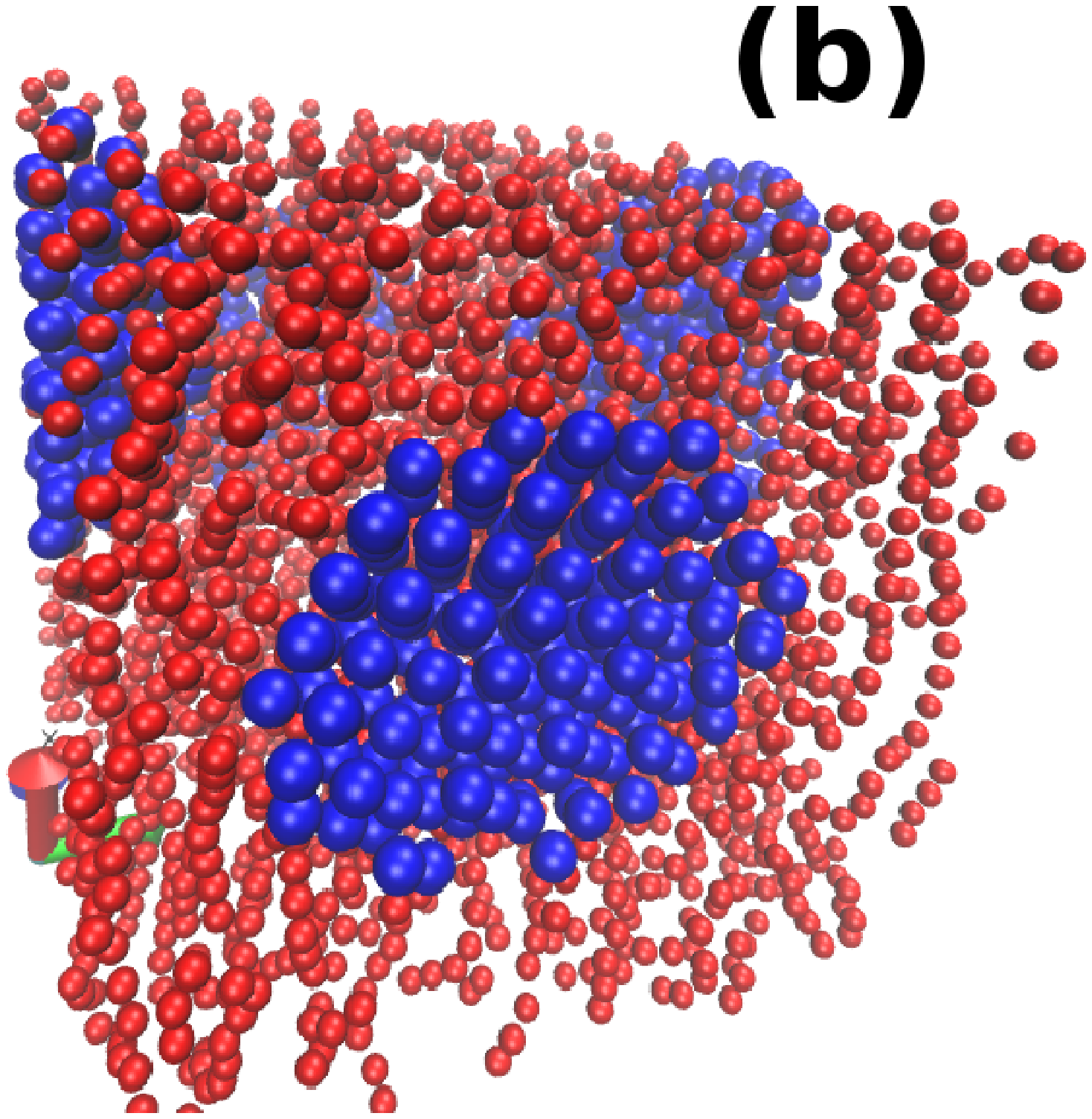}
\includegraphics[scale=0.3]{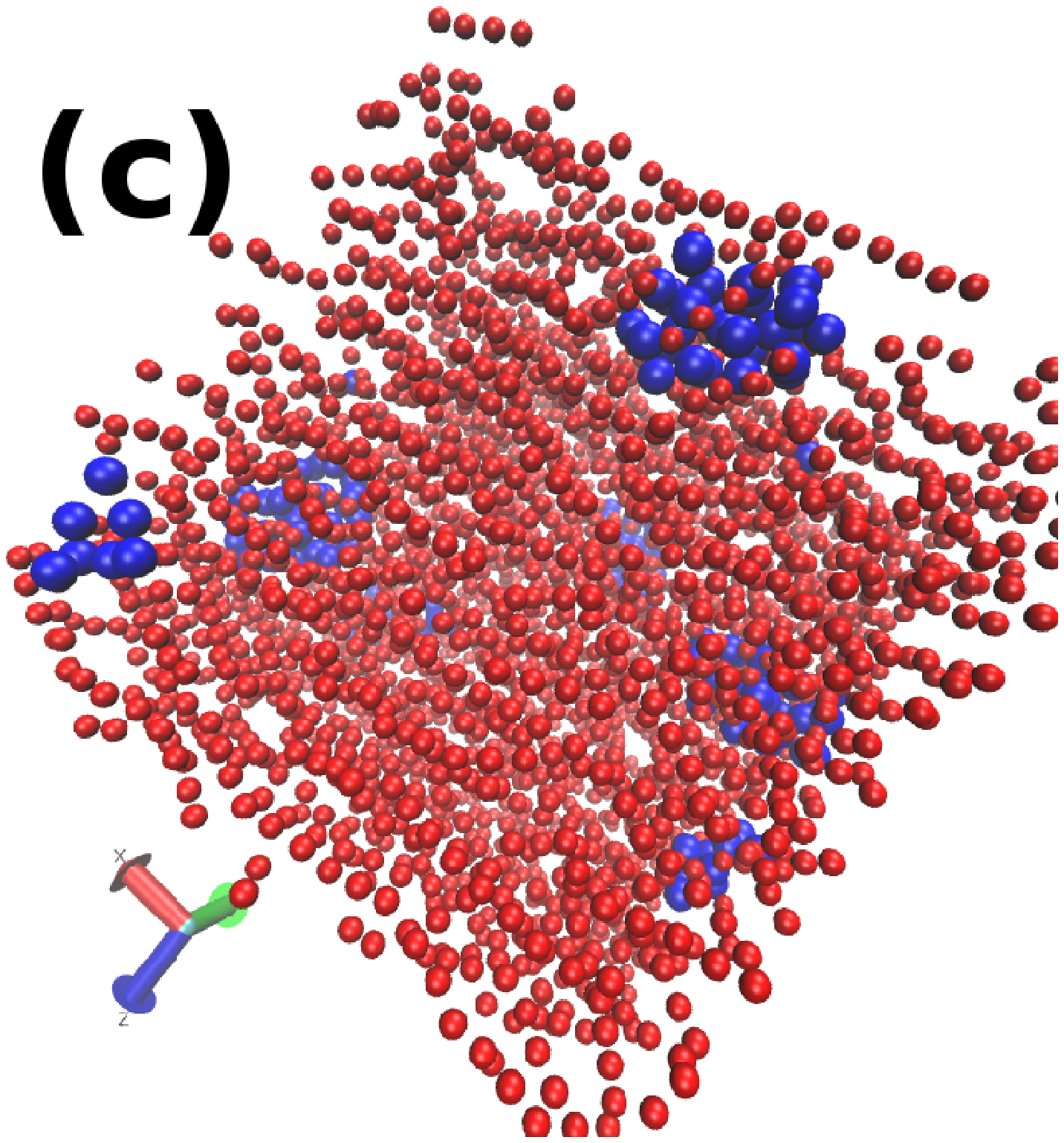}\\
\includegraphics[scale=0.3]{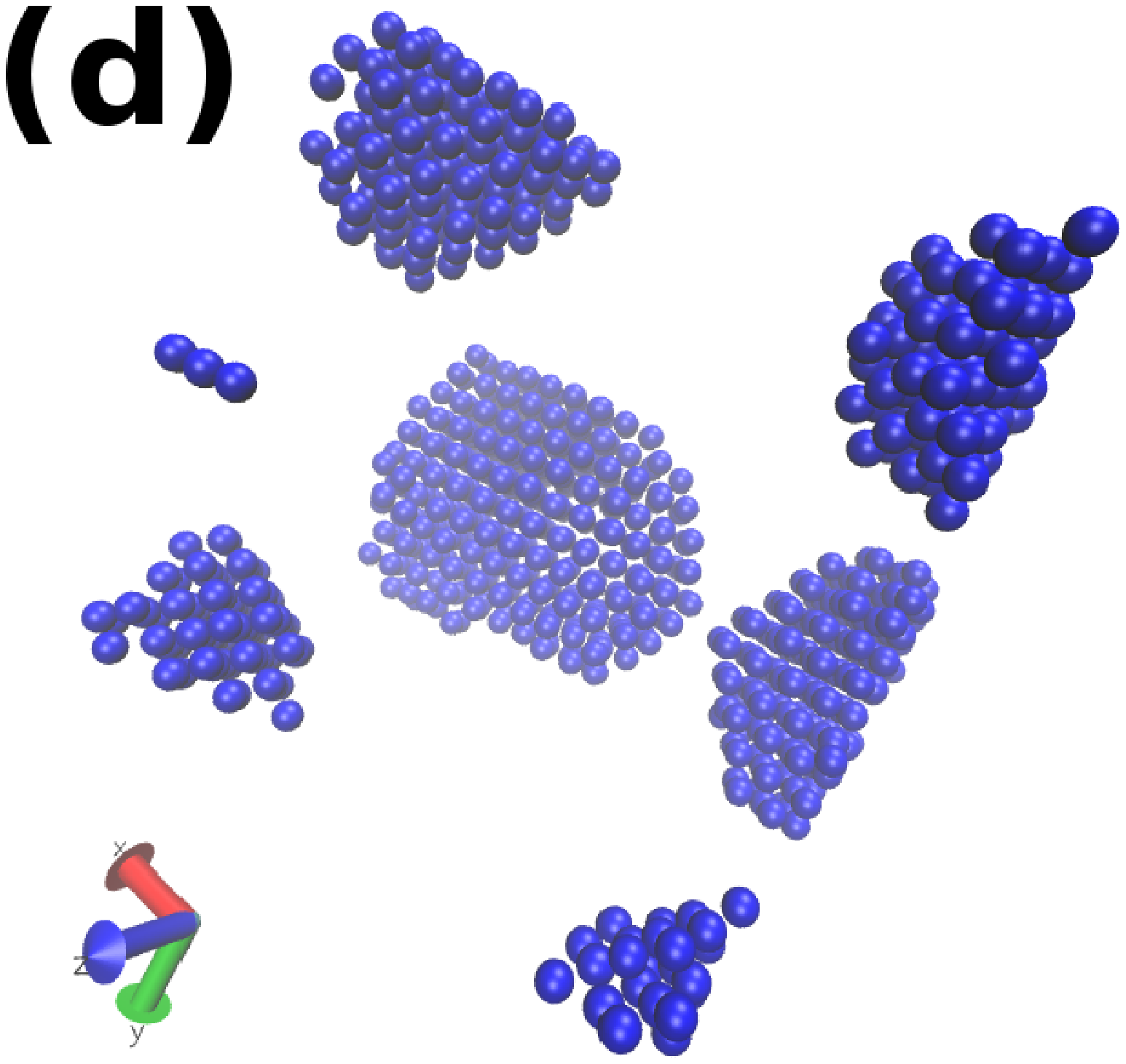}
\includegraphics[scale=0.3]{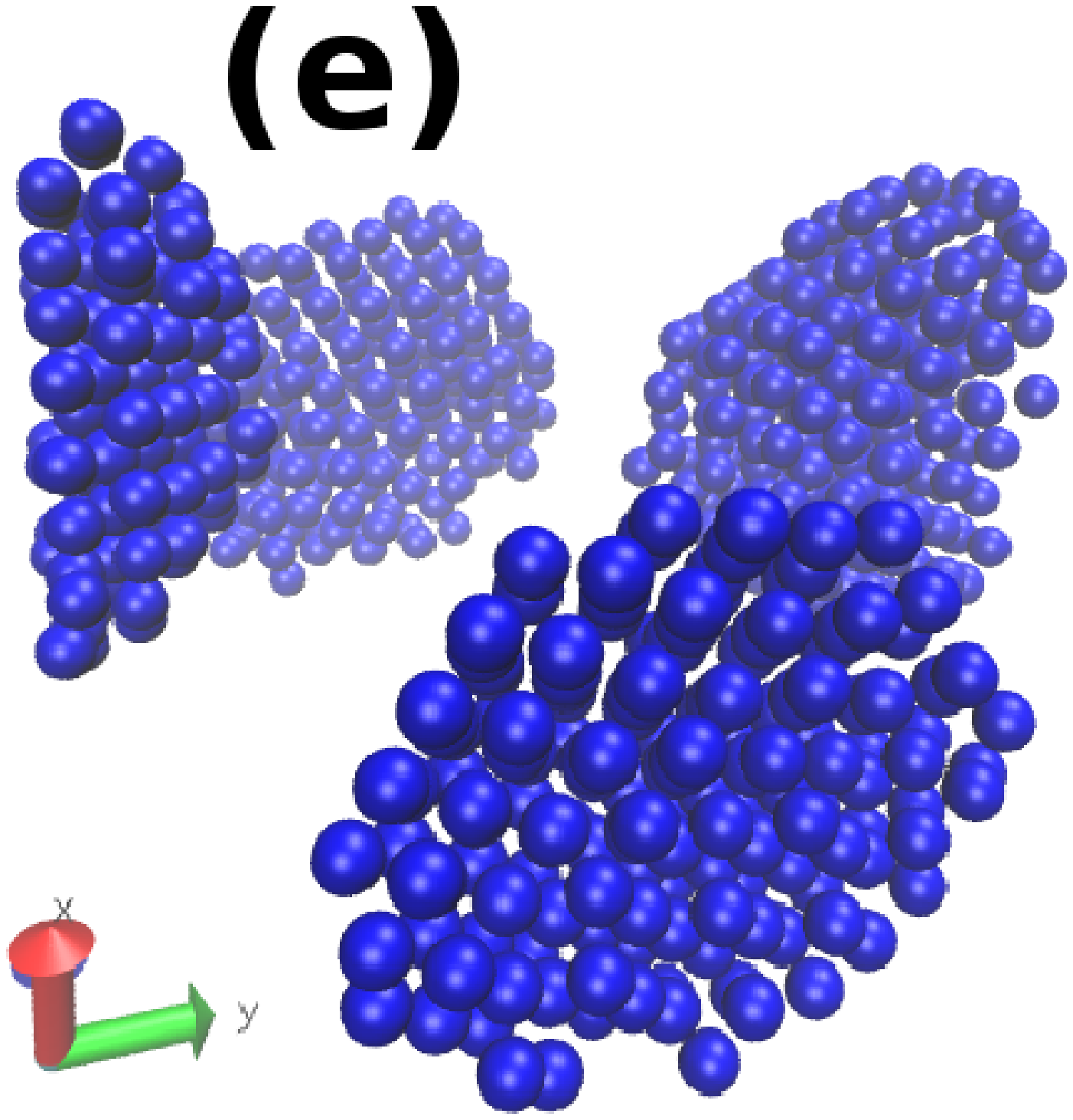}
\includegraphics[scale=0.3]{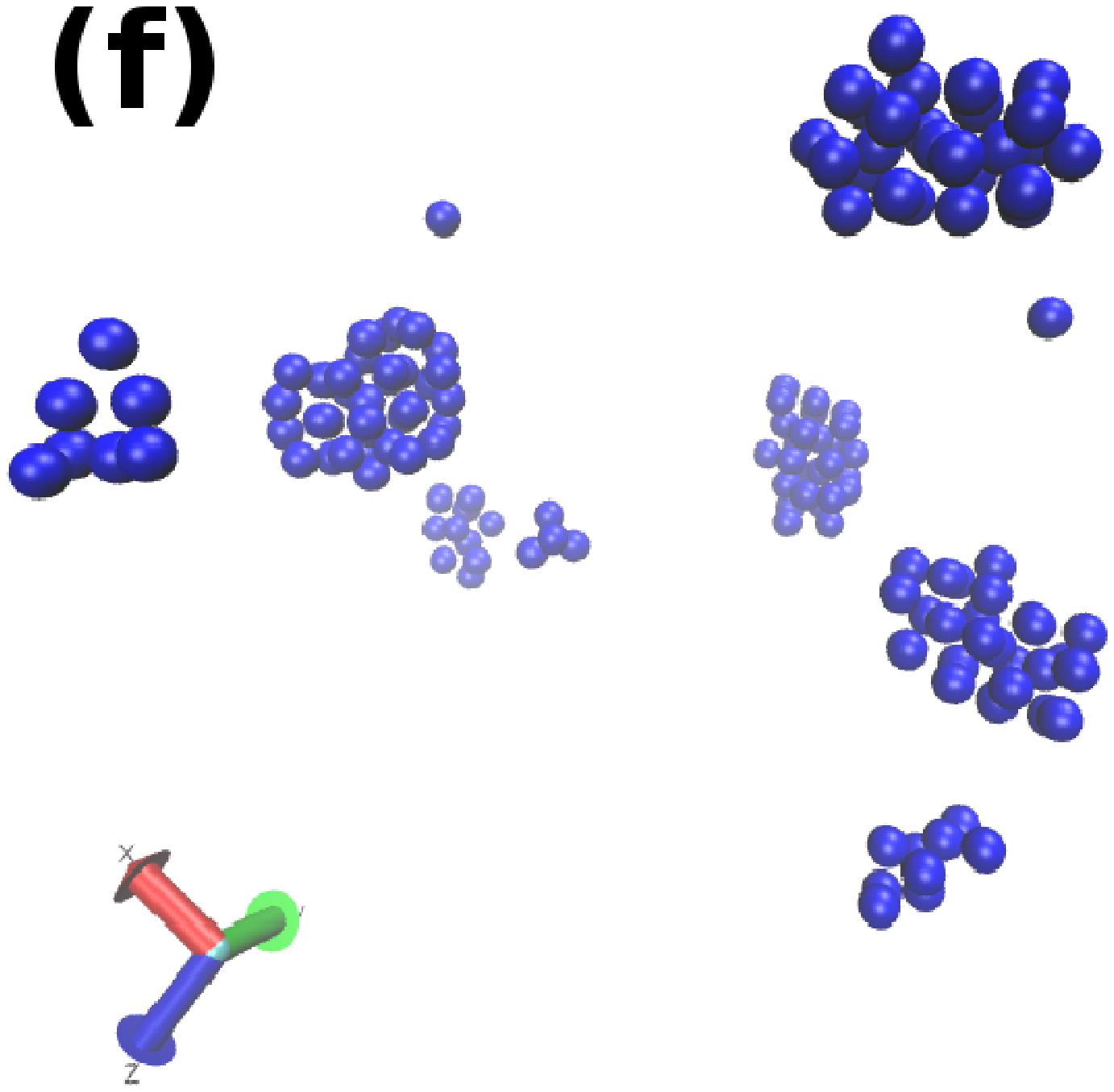}
\caption{The figure shows the snapshots of both nanoparticles and micelles (top row) and only the nanoparticles (bottom row) for different values of monomer densities (a) $0.089\sigma^{-3}$, (b) $0.1\sigma^{-3}$ and (c) $0.124\sigma^{-3}$ respectively. The size of nanoparticles and the shear rate are kept constant to be $\sigma_n=1.65$ and shear rate=$0.03\tau^{-1}$. All the snapshots of nanoparticles show a crystalline phase. However, due to the high density of the micellar matrix too few particles are able to introduce in the system to decode any order in the structure.}
\label{energies}
\end{figure*}

\begin{figure}
\centering
\includegraphics[scale=0.3]{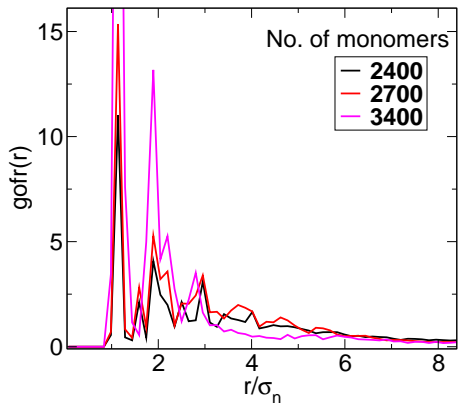}
\caption{ The figure shows the nanoparticle correlation function for the snapshots of nanoparticles shown in the lower row of Figure.5 for different values of monomer densities. The correlation function for both monomer densities of $\rho_m=0.089\sigma^{-3}$ and $0.1\sigma^{-3}$ show the existence of a long-range ordered phase while there are too few particles to decode a phase for the case of $\rho_m=0.124\sigma^{-3}$.}
\label{energies}
\end{figure}

\begin{figure*}
\centering
\includegraphics[scale=0.25]{0p03_2400_165_both_1.eps}
\includegraphics[scale=0.25]{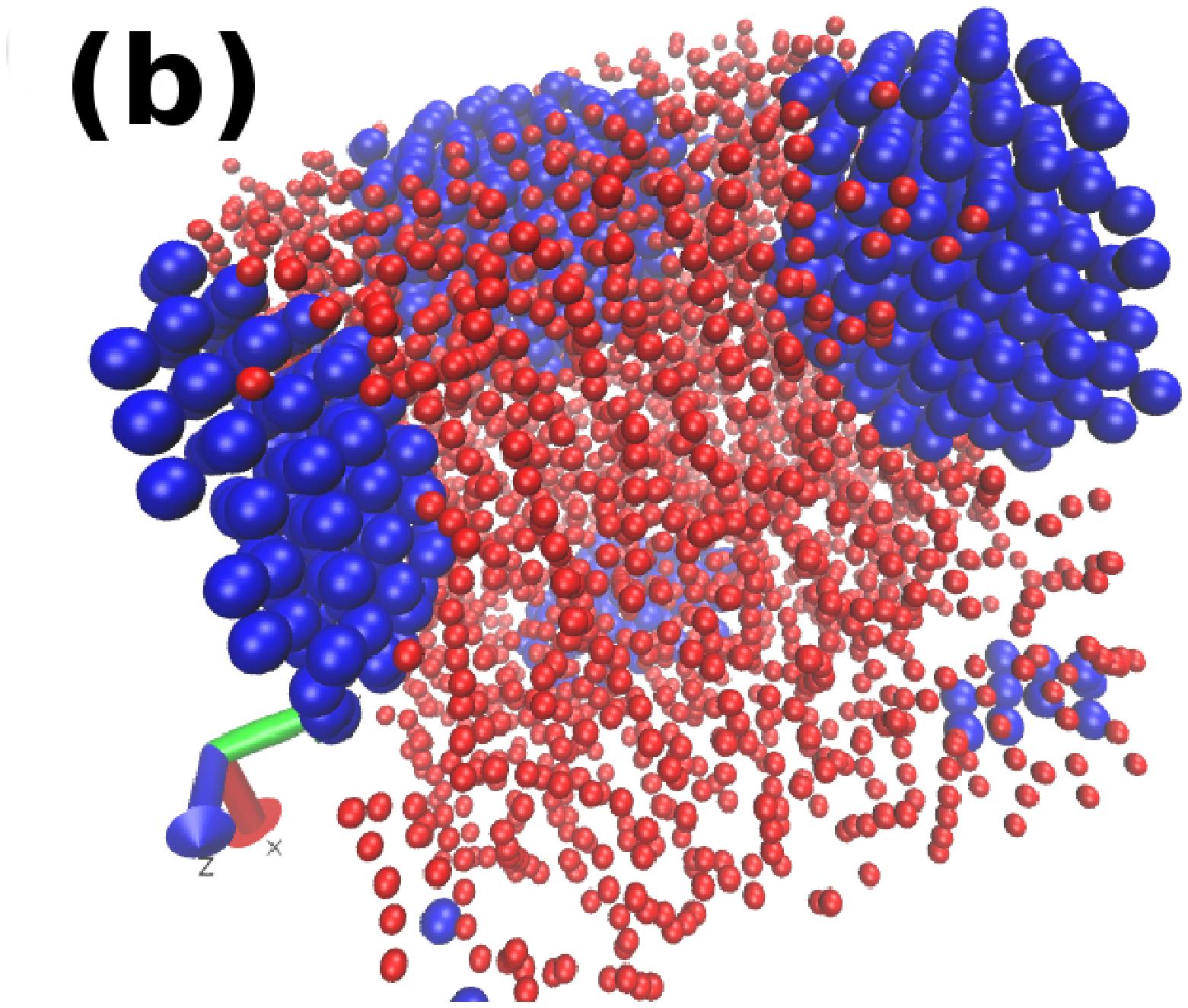}
\includegraphics[scale=0.25]{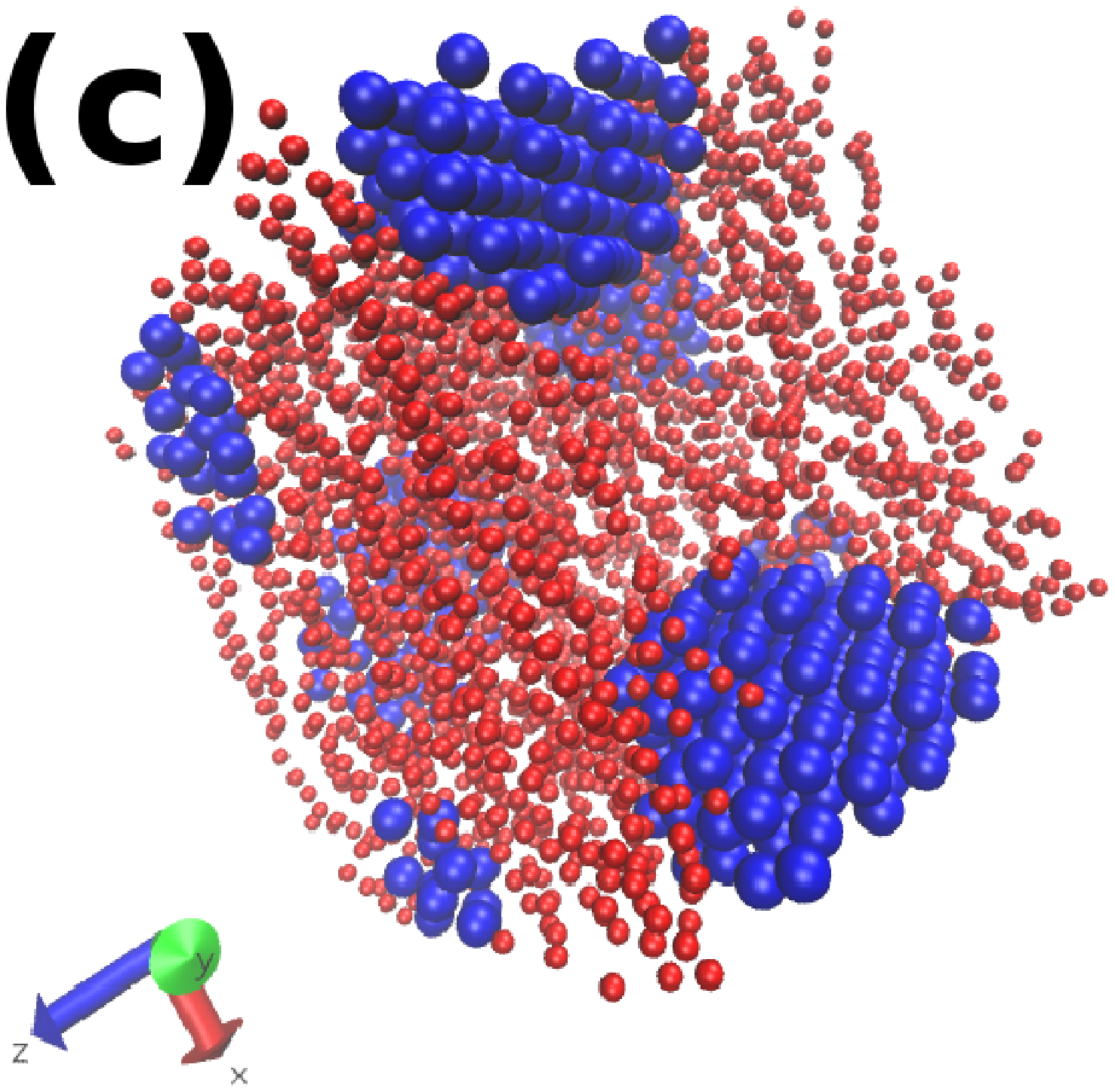}
\includegraphics[scale=0.25]{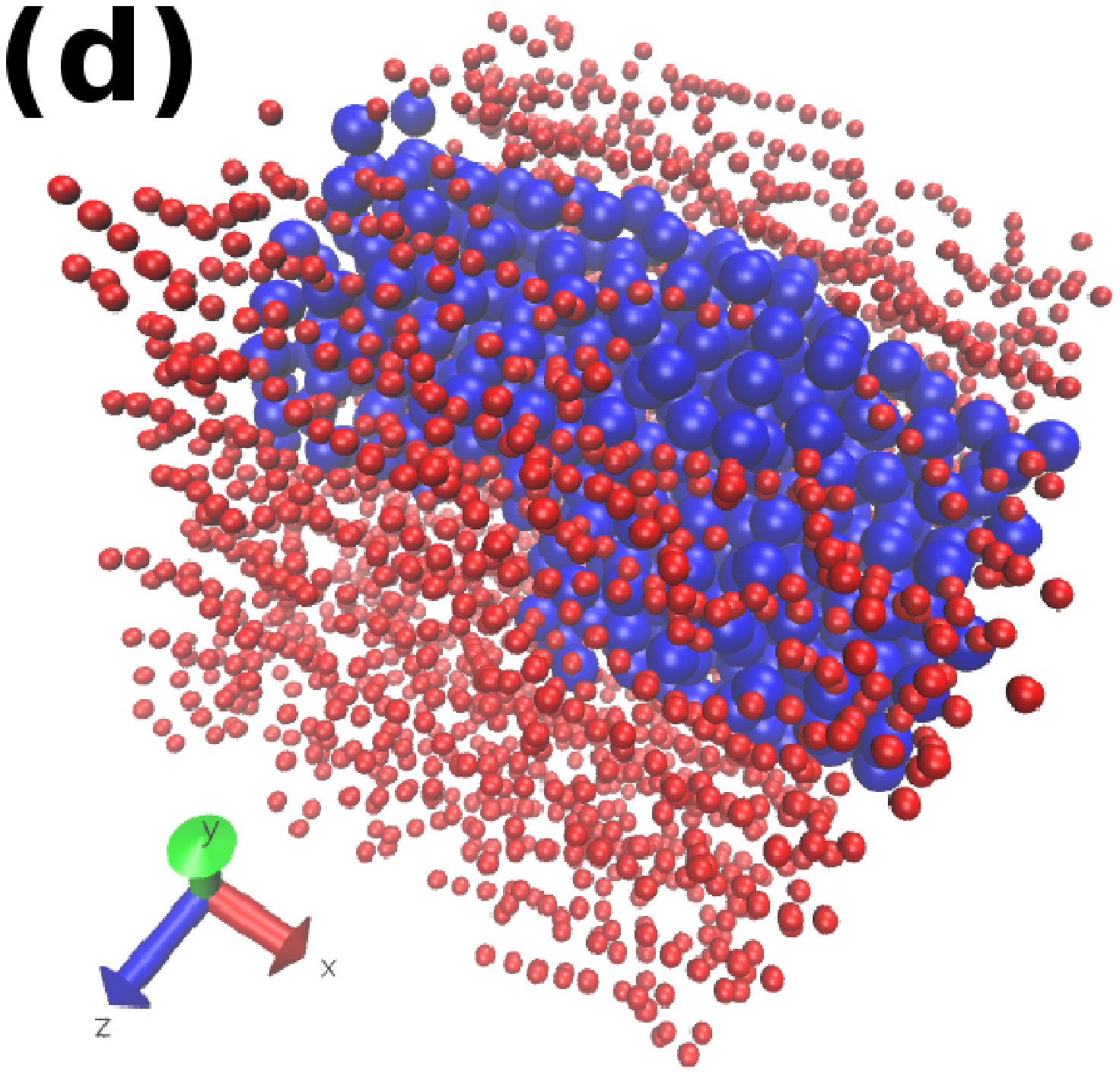}
\includegraphics[scale=0.25]{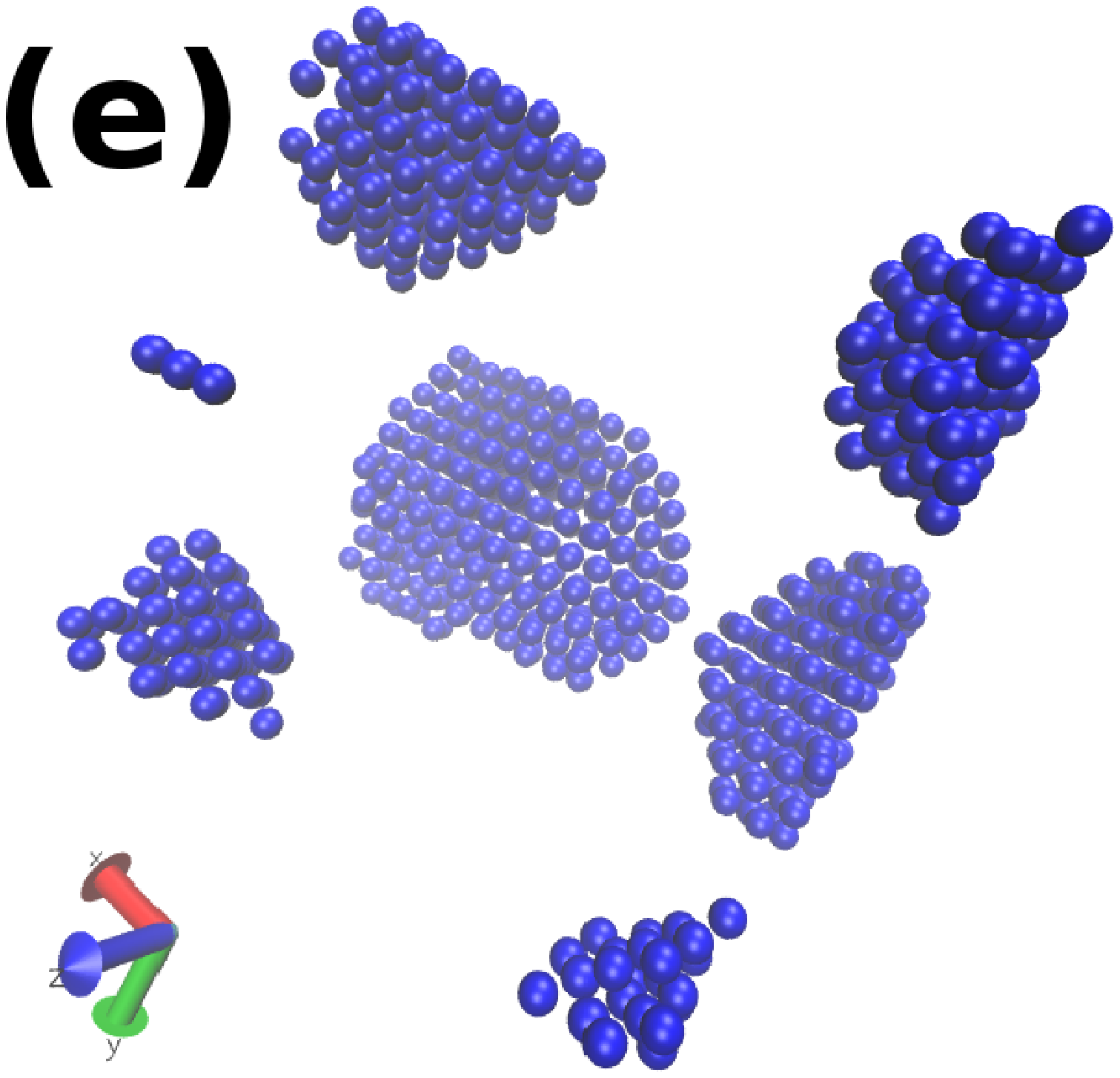}
\includegraphics[scale=0.25]{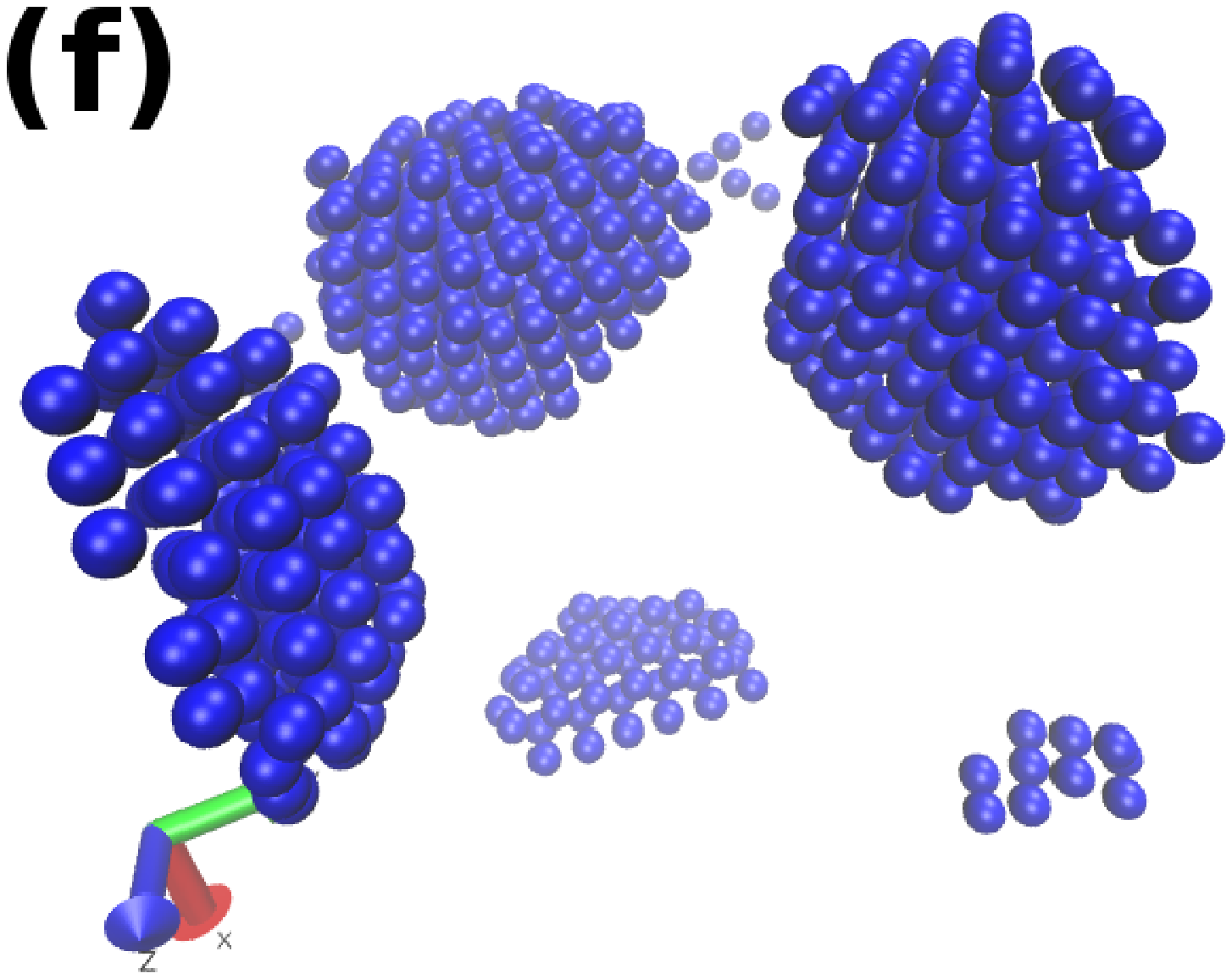}
\includegraphics[scale=0.25]{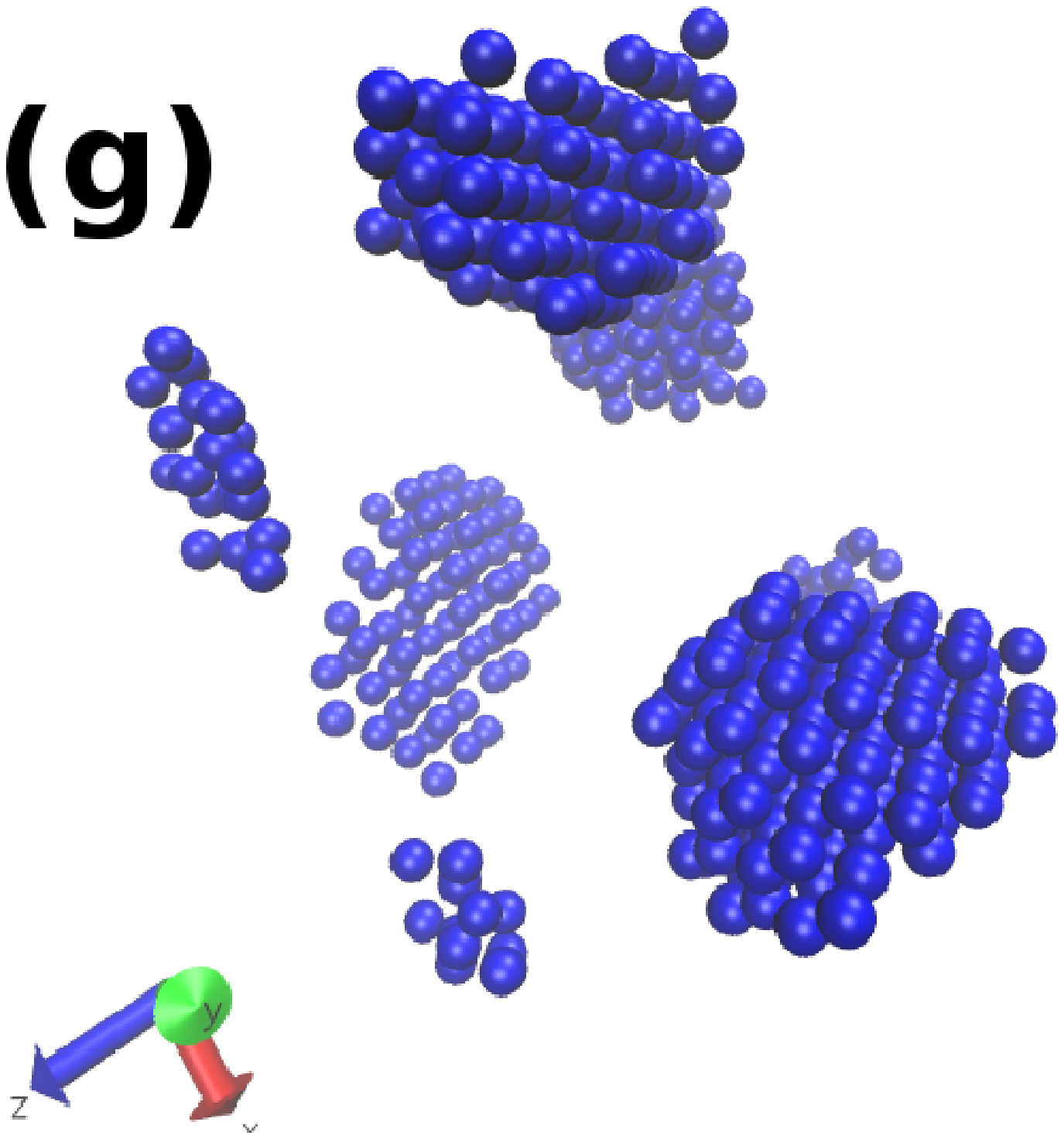}
\includegraphics[scale=0.25]{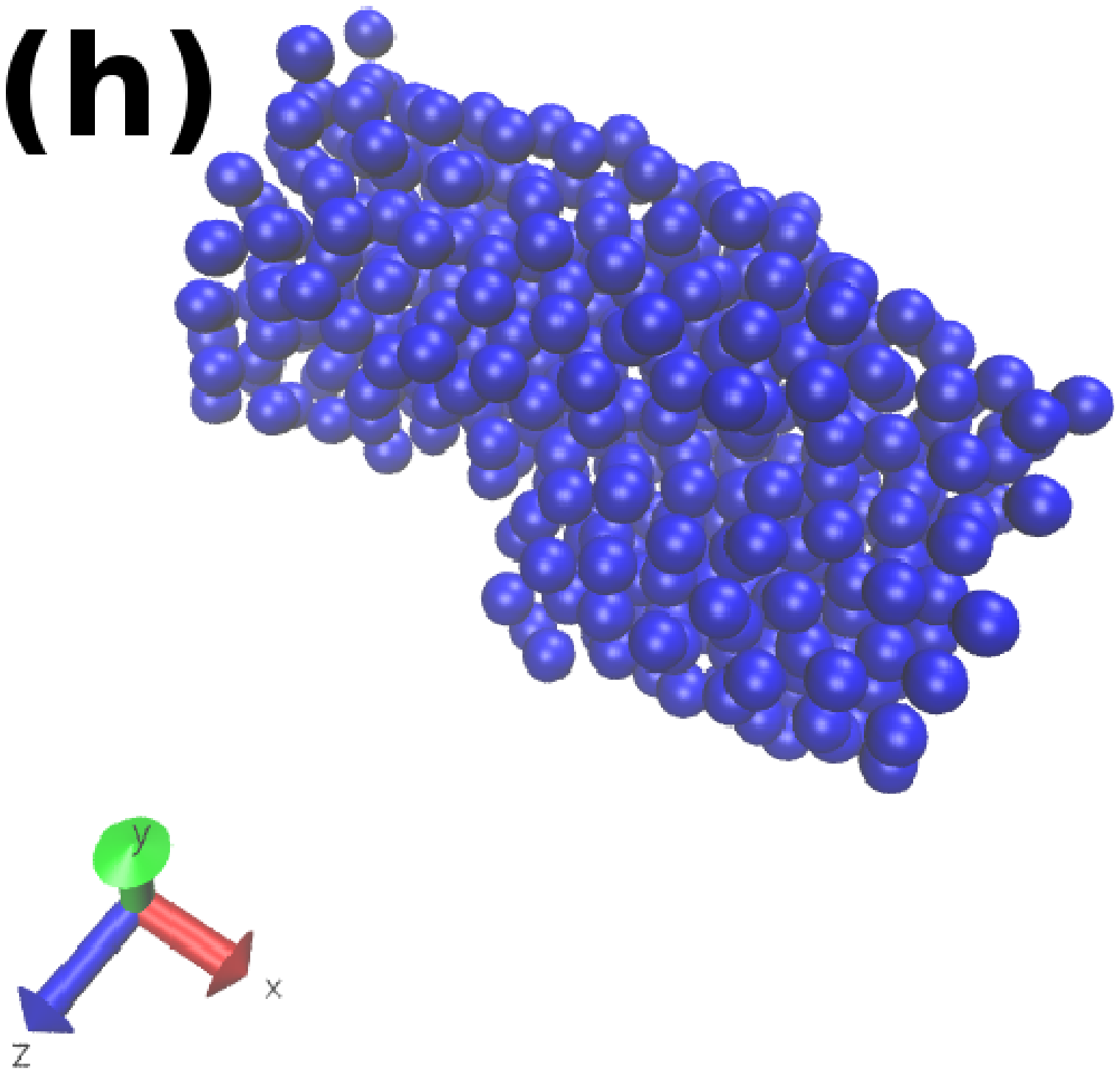}
\caption{ The figure shows the snapshots for four different values of nanoparticle size (a) $\sigma_n=1.65\sigma$, (b) $1.85\sigma$, (c) $2.05\sigma$ and (d) $2.25\sigma$. The micellar density is kept constant to $\rho_m=0.089\sigma^{-3}$ and the value of shear rate is kept fixed at $0.03\tau^{-1}$  The upper row indicates both the nanoparticles and monomers while, the lower row shows only nanoparticles. It can be clearly seen that the nanoparticle structure after crossing a particular value of $\sigma_n$ form a phase separated structure with a lower order of NP arrangement.}
\label{energies}
\end{figure*}

 The method of MPCD is first checked with only fluid particles in the simulation box and applying walls along the x-z planes. A bounce-back condition is maintained for the particles colliding with the wall and the phantom particles in the walls are taken into account. This generates a Poiseuille flow profile with a maximum velocity at the centre of the box and with no wall slip ~\cite{arxive4}.
Now, the walls along x-z planes are replaced by the shearing walls by imposing Lees-Edwards boundary conditions along the walls. This generates a shear flow~\cite{arxive4} with the velocity varying in the y-direction.
After checking the working of the MPCD method, monomers and nanoparticles are included in the box and the molecular dynamics simulation of the polymer-nanoparticle system is coupled with the MPCD fluid by including the monomers and nanoparticles in the collision step.
We denote the monomer number density as $\rho_m$.

In order to check the effect of shear rate on the WLM-Np system, a system of WLMs with $\rho_m=0.089\sigma^{-3}$ is considered. In the absence of shear, this system forms a percolating network of NPs ~\cite{arxive1}. Now the system is subject to shear with varying shear rates. It has already been shown that this system without NPs when subject to shear, first show an increase in its orientation and chain length and then a decrease in its chain length with the same orientation, as a result of increasing shear rate ~\cite{arxive4}. Similarly in the current system with NPs when the shear rate is increased from $0.001\tau^{-1}$ to $0.01\tau^{-1}$, the micellar chains show an increase in their chain length and orientation as shown in Fig. 1(a) and 1(b). In Fig. 1(a) micellar chains are shorter and have a low orientational order for shear rate=$0.001\tau^{-1}$. The corresponding nanostructure shows a percolating network in 1(d). Thus the NP morphology with this shear rate is hardly affected. However, increasing the shear rate to $0.1\tau^{-1}$ induces higher orientational order and length in the micellar chains with the NPs forming a single block of a long-range ordered particles (fig. 1(d)). This is a shear-induced phase separation of NPs. With further increase in shear rate to $0.03\tau^{-1}$, the NP structure breaks into smaller clusters.

       It is observed that the order of NP arrangement is highest in case of the intermediate value of the shear rate ($0.01\tau^{-1}$), when NPs form a phase-separated structure. This is confirmed by plotting the pair correlation function of NPs which is shown in Fig. 2. The pair correlation function shows the highest peak for shear rate=$0.01\tau^{-1}$. Moreover, the peaks around 2.6 and 3.55 indicate the hexagonal packing of the structures. Thus a well ordered crystalline structure of NPs can be formed by tuning the shear rate and the size of the NP crystalline block can also be controlled (refer Fig. 1e and 1f).

        The density of WLMs considered in Fig.1 has already been shown to produce sheet-like non-percolating nanostructures as can be seen in Fig. 1(f) ~\cite{arxive1}. However, the sheet-like non-percolating structures were produced as a result of an increase in the value of $\sigma_{4n}$ while in the case of Fig.1 they are produced as a result of shear. Now, to check what happens to the arrangement of NPs in these sheet-like non-percolating nanostructures when the micellar density $\rho_m$ is changed, the system considered in Fig.1(c) (or Fig.1(f)) is subject to the change in $\rho_m$. Since the shape anisotropy of the NP structure is controlled by the micellar density, we would like to remain in the regime where the micellar density produces only sheet-like nanostructures. The density $\rho_m=0.126\sigma^{-3}$ produces rod-like nanostructures ~\cite{arxive1}, therefore the density considered is below this value. The results are shown in Fig. 3.

         In Fig. 3, three different values of micellar densities are considered (a) $0.089\sigma^{-3}$, (b) $0.1\sigma^{-3}$ and (c) $0.124\sigma^{-3}$. As the micellar density increases, it becomes very difficult to introduce NPs inside the matrix and shear it. Therefore, for the highest density considered, hardly few NPs are able to introduce (Fig. 3(c)). Now, the system is sheared with the shear rate=$0.03\sigma^{-3}$ keeping the size of nanoparticles same as $\sigma_n=1.65\sigma$. From the figure, we can clearly see that both figures 3(d) and 3(f) maintains the sheet-like morphology of NPs with the long-range ordered arrangement of particles. However, the shear rate seems to affect the crystalline order of the NP structures. This is confirmed by plotting the pair correlation function of the NPs structures which is shown in Fig. 4. In this figure, the pair correlation function shows higher peaks for the NPs in the system with $\rho_m=0.1\sigma^{-3}$ (Fig 3(e)) compared to the structure in case of $\rho_m=0.089\sigma^{-3}$ (Fig.1(d)). Thus, increasing the density of the micellar matrix increases the crystalline order of the NP structure.

\begin{figure}
\centering
\includegraphics[scale=0.3]{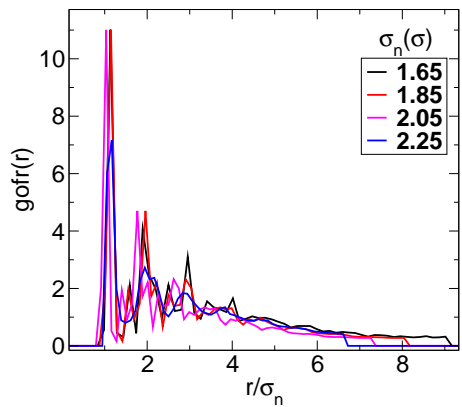}
\caption{ The figure shows the pair correlation functions for four different values of nanoparticle size whose snapshots are shown in the lower row of figure.7. The pair correlation functions for smaller size particles show a long-range ordered structures with well-defined peaks while the correlation function for the largest particle considered shows a lower packing of the structure with broad and lower peaks.}
\label{energies}
\end{figure}

    It is also insightful to investigate the effect of the size of NPs when the WLM-NP system is sheared. In our earlier investigation, it is shown that the NPs with smaller size are capable of forming a well packed-structure. Increasing the size of NPs affects the long-range order of the nanostructure and sometimes not a well-defined structure is observed ~\cite{arxive2}.  Now to investigate the effect of the size of NPs on the WLM-NP system when it is sheared, the same system shown in Fig. 3(a) (or 1(c)) is considered and the value of $\sigma_n$ is varied with the shear rate kept fixed at $0.03\tau^{-1}$. The results are shown in Fig. 5. The figure shows the system with $\sigma_n$ (a) $1.65\sigma$, (b) $1.85\sigma$, (c) $2.05\sigma$ and (d) $2.25\sigma$. Examining the figure, we can see that with increasing the size of NPs, the shape anisotropy of the NP structure is getting affected with the NPs showing a phase separated structure after crossing a particular value of $\sigma_n$. This can be seen in Fig. 5(h) (or Fig. 5(d)), where the NPs form a phase-separated structure in the matrix. Moreover, the size of NPs also seems to affect the arrangement of NPs.

       The arrangement of the NP structure can be analyzed by plotting their pair correlation function. The plot is shown in Fig. 6. In the figure, it can be seen that the pair correlation functions for $\sigma_n=1.65\sigma$, $1.85\sigma$ and $2.05\sigma$ (when the NPs form non-percolating sheet-like structure (Fig. 5(e), (f), and (g) )), shows higher peaks indicating the presence of a long-range order in the structure. However, the peaks for $\sigma_n=2.25\sigma$ is very low that shows not a well-ordered structure which is also confirmed from Fig. 5(h).
The phase separation of the NP structure for the higher size of NPs ($\sigma_n=2.25\sigma$) is due to the surface energy of the NPs. With the increase in the size of NPs, the energy of NPs in a cluster decreases, hence NPs are unable to withstand in the shear if they form a small cluster. Therefore, it is more cost effective for the NPs to form a big cluster and phase separate in order to minimize surface interaction with the shearing micellar matrix. Since the order of NP arrangement is an important factor affecting the mechanical properties of elastomers or flexible materials, it is very important to consider the size of NPs in order to device wearable, flexible or photonic materials.

\bibliographystyle{apsrev4-1}
\bibliography{paper_shear_nano}

\begin{thebibliography}{34}%
\makeatletter
\providecommand \@ifxundefined [1]{%
 \@ifx{#1\undefined}
}%
\providecommand \@ifnum [1]{%
 \ifnum #1\expandafter \@firstoftwo
 \else \expandafter \@secondoftwo
 \fi
}%
\providecommand \@ifx [1]{%
 \ifx #1\expandafter \@firstoftwo
 \else \expandafter \@secondoftwo
 \fi
}%
\providecommand \natexlab [1]{#1}%
\providecommand \enquote  [1]{``#1''}%
\providecommand \bibnamefont  [1]{#1}%
\providecommand \bibfnamefont [1]{#1}%
\providecommand \citenamefont [1]{#1}%
\providecommand \href@noop [0]{\@secondoftwo}%
\providecommand \href [0]{\begingroup \@sanitize@url \@href}%
\providecommand \@href[1]{\@@startlink{#1}\@@href}%
\providecommand \@@href[1]{\endgroup#1\@@endlink}%
\providecommand \@sanitize@url [0]{\catcode `\\12\catcode `\$12\catcode
  `\&12\catcode `\#12\catcode `\^12\catcode `\_12\catcode `\%12\relax}%
\providecommand \@@startlink[1]{}%
\providecommand \@@endlink[0]{}%
\providecommand \url  [0]{\begingroup\@sanitize@url \@url }%
\providecommand \@url [1]{\endgroup\@href {#1}{\urlprefix }}%
\providecommand \urlprefix  [0]{URL }%
\providecommand \Eprint [0]{\href }%
\providecommand \doibase [0]{http://dx.doi.org/}%
\providecommand \selectlanguage [0]{\@gobble}%
\providecommand \bibinfo  [0]{\@secondoftwo}%
\providecommand \bibfield  [0]{\@secondoftwo}%
\providecommand \translation [1]{[#1]}%
\providecommand \BibitemOpen [0]{}%
\providecommand \bibitemStop [0]{}%
\providecommand \bibitemNoStop [0]{.\EOS\space}%
\providecommand \EOS [0]{\spacefactor3000\relax}%
\providecommand \BibitemShut  [1]{\csname bibitem#1\endcsname}%
\let\auto@bib@innerbib\@empty
%</preamble>
\bibitem [{\citenamefont {Nakamura}\ \emph {et~al.}(1992)\citenamefont
  {Nakamura}, \citenamefont {Yamaguchi}, \citenamefont {Okubo},\ and\
  \citenamefont {Matsumoto}}]{nakamura1992effect}%
  \BibitemOpen
  \bibfield  {author} {\bibinfo {author} {\bibfnamefont {Y.}~\bibnamefont
  {Nakamura}}, \bibinfo {author} {\bibfnamefont {M.}~\bibnamefont {Yamaguchi}},
  \bibinfo {author} {\bibfnamefont {M.}~\bibnamefont {Okubo}}, \ and\ \bibinfo
  {author} {\bibfnamefont {T.}~\bibnamefont {Matsumoto}},\ }\href@noop {}
  {\bibfield  {journal} {\bibinfo  {journal} {Journal of applied polymer
  science}\ }\textbf {\bibinfo {volume} {44}},\ \bibinfo {pages} {151}
  (\bibinfo {year} {1992})}\BibitemShut {NoStop}%
\bibitem [{\citenamefont {Moloney}\ \emph {et~al.}(1984)\citenamefont
  {Moloney}, \citenamefont {Kausch},\ and\ \citenamefont
  {Stieger}}]{moloney1984fracture}%
  \BibitemOpen
  \bibfield  {author} {\bibinfo {author} {\bibfnamefont {A.}~\bibnamefont
  {Moloney}}, \bibinfo {author} {\bibfnamefont {H.}~\bibnamefont {Kausch}}, \
  and\ \bibinfo {author} {\bibfnamefont {H.}~\bibnamefont {Stieger}},\
  }\href@noop {} {\bibfield  {journal} {\bibinfo  {journal} {Journal of
  materials science}\ }\textbf {\bibinfo {volume} {19}},\ \bibinfo {pages}
  {1125} (\bibinfo {year} {1984})}\BibitemShut {NoStop}%
\bibitem [{\citenamefont {Leblanc}(2002)}]{leblanc2002rubber}%
  \BibitemOpen
  \bibfield  {author} {\bibinfo {author} {\bibfnamefont {J.~L.}\ \bibnamefont
  {Leblanc}},\ }\href@noop {} {\bibfield  {journal} {\bibinfo  {journal}
  {Progress in polymer science}\ }\textbf {\bibinfo {volume} {27}},\ \bibinfo
  {pages} {627} (\bibinfo {year} {2002})}\BibitemShut {NoStop}%
\bibitem [{\citenamefont {Braun}\ \emph {et~al.}(2006)\citenamefont {Braun},
  \citenamefont {Rinne},\ and\ \citenamefont
  {Garc{\'\i}a-Santamar{\'\i}a}}]{braun2006introducing}%
  \BibitemOpen
  \bibfield  {author} {\bibinfo {author} {\bibfnamefont {P.~V.}\ \bibnamefont
  {Braun}}, \bibinfo {author} {\bibfnamefont {S.~A.}\ \bibnamefont {Rinne}}, \
  and\ \bibinfo {author} {\bibfnamefont {F.}~\bibnamefont
  {Garc{\'\i}a-Santamar{\'\i}a}},\ }\href@noop {} {\bibfield  {journal}
  {\bibinfo  {journal} {Advanced Materials}\ }\textbf {\bibinfo {volume}
  {18}},\ \bibinfo {pages} {2665} (\bibinfo {year} {2006})}\BibitemShut
  {NoStop}%
\bibitem [{\citenamefont {Van~Blaaderen}(1998)}]{van1998opals}%
  \BibitemOpen
  \bibfield  {author} {\bibinfo {author} {\bibfnamefont {A.}~\bibnamefont
  {Van~Blaaderen}},\ }\href@noop {} {\bibfield  {journal} {\bibinfo  {journal}
  {Science}\ }\textbf {\bibinfo {volume} {282}},\ \bibinfo {pages} {887}
  (\bibinfo {year} {1998})}\BibitemShut {NoStop}%
\bibitem [{\citenamefont {Gates}(2009)}]{gates2009flexible}%
  \BibitemOpen
  \bibfield  {author} {\bibinfo {author} {\bibfnamefont {B.~D.}\ \bibnamefont
  {Gates}},\ }\href@noop {} {\bibfield  {journal} {\bibinfo  {journal}
  {Science}\ }\textbf {\bibinfo {volume} {323}},\ \bibinfo {pages} {1566}
  (\bibinfo {year} {2009})}\BibitemShut {NoStop}%
\bibitem [{\citenamefont {Kim}\ \emph {et~al.}(2012)\citenamefont {Kim},
  \citenamefont {Kim}, \citenamefont {Kwon}, \citenamefont {Kim}, \citenamefont
  {Park}, \citenamefont {Yoon}, \citenamefont {Jang}, \citenamefont {Shin},
  \citenamefont {Suh},\ and\ \citenamefont {Yang}}]{kim2012large}%
  \BibitemOpen
  \bibfield  {author} {\bibinfo {author} {\bibfnamefont {T.}~\bibnamefont
  {Kim}}, \bibinfo {author} {\bibfnamefont {H.}~\bibnamefont {Kim}}, \bibinfo
  {author} {\bibfnamefont {S.~W.}\ \bibnamefont {Kwon}}, \bibinfo {author}
  {\bibfnamefont {Y.}~\bibnamefont {Kim}}, \bibinfo {author} {\bibfnamefont
  {W.~K.}\ \bibnamefont {Park}}, \bibinfo {author} {\bibfnamefont {D.~H.}\
  \bibnamefont {Yoon}}, \bibinfo {author} {\bibfnamefont {A.-R.}\ \bibnamefont
  {Jang}}, \bibinfo {author} {\bibfnamefont {H.~S.}\ \bibnamefont {Shin}},
  \bibinfo {author} {\bibfnamefont {K.~S.}\ \bibnamefont {Suh}}, \ and\
  \bibinfo {author} {\bibfnamefont {W.~S.}\ \bibnamefont {Yang}},\ }\href@noop
  {} {\bibfield  {journal} {\bibinfo  {journal} {Nano letters}\ }\textbf
  {\bibinfo {volume} {12}},\ \bibinfo {pages} {743} (\bibinfo {year}
  {2012})}\BibitemShut {NoStop}%
\bibitem [{\citenamefont {Mannsfeld}\ \emph {et~al.}(2010)\citenamefont
  {Mannsfeld}, \citenamefont {Tee}, \citenamefont {Stoltenberg}, \citenamefont
  {Chen}, \citenamefont {Barman}, \citenamefont {Muir}, \citenamefont
  {Sokolov}, \citenamefont {Reese},\ and\ \citenamefont
  {Bao}}]{mannsfeld2010highly}%
  \BibitemOpen
  \bibfield  {author} {\bibinfo {author} {\bibfnamefont {S.~C.}\ \bibnamefont
  {Mannsfeld}}, \bibinfo {author} {\bibfnamefont {B.~C.}\ \bibnamefont {Tee}},
  \bibinfo {author} {\bibfnamefont {R.~M.}\ \bibnamefont {Stoltenberg}},
  \bibinfo {author} {\bibfnamefont {C.~V.~H.}\ \bibnamefont {Chen}}, \bibinfo
  {author} {\bibfnamefont {S.}~\bibnamefont {Barman}}, \bibinfo {author}
  {\bibfnamefont {B.~V.}\ \bibnamefont {Muir}}, \bibinfo {author}
  {\bibfnamefont {A.~N.}\ \bibnamefont {Sokolov}}, \bibinfo {author}
  {\bibfnamefont {C.}~\bibnamefont {Reese}}, \ and\ \bibinfo {author}
  {\bibfnamefont {Z.}~\bibnamefont {Bao}},\ }\href@noop {} {\bibfield
  {journal} {\bibinfo  {journal} {Nature materials}\ }\textbf {\bibinfo
  {volume} {9}},\ \bibinfo {pages} {859} (\bibinfo {year} {2010})}\BibitemShut
  {NoStop}%
\bibitem [{\citenamefont {Chen}\ \emph {et~al.}(1994)\citenamefont {Chen},
  \citenamefont {Ackerson},\ and\ \citenamefont
  {Zukoski}}]{chen1994rheological}%
  \BibitemOpen
  \bibfield  {author} {\bibinfo {author} {\bibfnamefont {L.}~\bibnamefont
  {Chen}}, \bibinfo {author} {\bibfnamefont {B.}~\bibnamefont {Ackerson}}, \
  and\ \bibinfo {author} {\bibfnamefont {C.}~\bibnamefont {Zukoski}},\
  }\href@noop {} {\bibfield  {journal} {\bibinfo  {journal} {Journal of
  Rheology}\ }\textbf {\bibinfo {volume} {38}},\ \bibinfo {pages} {193}
  (\bibinfo {year} {1994})}\BibitemShut {NoStop}%
\bibitem [{\citenamefont {Ciamarra}\ \emph {et~al.}(2005)\citenamefont
  {Ciamarra}, \citenamefont {Coniglio},\ and\ \citenamefont
  {Nicodemi}}]{ciamarra2005shear}%
  \BibitemOpen
  \bibfield  {author} {\bibinfo {author} {\bibfnamefont {M.~P.}\ \bibnamefont
  {Ciamarra}}, \bibinfo {author} {\bibfnamefont {A.}~\bibnamefont {Coniglio}},
  \ and\ \bibinfo {author} {\bibfnamefont {M.}~\bibnamefont {Nicodemi}},\
  }\href@noop {} {\bibfield  {journal} {\bibinfo  {journal} {Physical review
  letters}\ }\textbf {\bibinfo {volume} {94}},\ \bibinfo {pages} {188001}
  (\bibinfo {year} {2005})}\BibitemShut {NoStop}%
\bibitem [{\citenamefont {Fan}\ and\ \citenamefont
  {Hill}(2011)}]{fan2011phase}%
  \BibitemOpen
  \bibfield  {author} {\bibinfo {author} {\bibfnamefont {Y.}~\bibnamefont
  {Fan}}\ and\ \bibinfo {author} {\bibfnamefont {K.}~\bibnamefont {Hill}},\
  }\href@noop {} {\bibfield  {journal} {\bibinfo  {journal} {Physical review
  letters}\ }\textbf {\bibinfo {volume} {106}},\ \bibinfo {pages} {218301}
  (\bibinfo {year} {2011})}\BibitemShut {NoStop}%
\bibitem [{\citenamefont {Fernandez}\ \emph {et~al.}(2013)\citenamefont
  {Fernandez}, \citenamefont {Mani}, \citenamefont {Rinaldi}, \citenamefont
  {Kadau}, \citenamefont {Mosquet}, \citenamefont {Lombois-Burger},
  \citenamefont {Cayer-Barrioz}, \citenamefont {Herrmann}, \citenamefont
  {Spencer},\ and\ \citenamefont {Isa}}]{fernandez2013microscopic}%
  \BibitemOpen
  \bibfield  {author} {\bibinfo {author} {\bibfnamefont {N.}~\bibnamefont
  {Fernandez}}, \bibinfo {author} {\bibfnamefont {R.}~\bibnamefont {Mani}},
  \bibinfo {author} {\bibfnamefont {D.}~\bibnamefont {Rinaldi}}, \bibinfo
  {author} {\bibfnamefont {D.}~\bibnamefont {Kadau}}, \bibinfo {author}
  {\bibfnamefont {M.}~\bibnamefont {Mosquet}}, \bibinfo {author} {\bibfnamefont
  {H.}~\bibnamefont {Lombois-Burger}}, \bibinfo {author} {\bibfnamefont
  {J.}~\bibnamefont {Cayer-Barrioz}}, \bibinfo {author} {\bibfnamefont {H.~J.}\
  \bibnamefont {Herrmann}}, \bibinfo {author} {\bibfnamefont {N.~D.}\
  \bibnamefont {Spencer}}, \ and\ \bibinfo {author} {\bibfnamefont
  {L.}~\bibnamefont {Isa}},\ }\href@noop {} {\bibfield  {journal} {\bibinfo
  {journal} {Physical review letters}\ }\textbf {\bibinfo {volume} {111}},\
  \bibinfo {pages} {108301} (\bibinfo {year} {2013})}\BibitemShut {NoStop}%
\bibitem [{\citenamefont {Besseling}\ \emph {et~al.}(2012)\citenamefont
  {Besseling}, \citenamefont {Hermes}, \citenamefont {Fortini}, \citenamefont
  {Dijkstra}, \citenamefont {Imhof},\ and\ \citenamefont
  {Van~Blaaderen}}]{besseling2012oscillatory}%
  \BibitemOpen
  \bibfield  {author} {\bibinfo {author} {\bibfnamefont {T.}~\bibnamefont
  {Besseling}}, \bibinfo {author} {\bibfnamefont {M.}~\bibnamefont {Hermes}},
  \bibinfo {author} {\bibfnamefont {A.}~\bibnamefont {Fortini}}, \bibinfo
  {author} {\bibfnamefont {M.}~\bibnamefont {Dijkstra}}, \bibinfo {author}
  {\bibfnamefont {A.}~\bibnamefont {Imhof}}, \ and\ \bibinfo {author}
  {\bibfnamefont {A.}~\bibnamefont {Van~Blaaderen}},\ }\href@noop {} {\bibfield
   {journal} {\bibinfo  {journal} {Soft Matter}\ }\textbf {\bibinfo {volume}
  {8}},\ \bibinfo {pages} {6931} (\bibinfo {year} {2012})}\BibitemShut
  {NoStop}%
\bibitem [{\citenamefont {Zheng}\ \emph {et~al.}(2016)\citenamefont {Zheng},
  \citenamefont {Wang}, \citenamefont {Wang}, \citenamefont {Liu},
  \citenamefont {Wu},\ and\ \citenamefont {Zhang}}]{zheng2016dispersion}%
  \BibitemOpen
  \bibfield  {author} {\bibinfo {author} {\bibfnamefont {Z.}~\bibnamefont
  {Zheng}}, \bibinfo {author} {\bibfnamefont {Z.}~\bibnamefont {Wang}},
  \bibinfo {author} {\bibfnamefont {L.}~\bibnamefont {Wang}}, \bibinfo {author}
  {\bibfnamefont {J.}~\bibnamefont {Liu}}, \bibinfo {author} {\bibfnamefont
  {Y.}~\bibnamefont {Wu}}, \ and\ \bibinfo {author} {\bibfnamefont
  {L.}~\bibnamefont {Zhang}},\ }\href@noop {} {\bibfield  {journal} {\bibinfo
  {journal} {Nanotechnology}\ }\textbf {\bibinfo {volume} {27}},\ \bibinfo
  {pages} {265704} (\bibinfo {year} {2016})}\BibitemShut {NoStop}%
\bibitem [{\citenamefont {Park}\ \emph {et~al.}(2014)\citenamefont {Park},
  \citenamefont {Kalra},\ and\ \citenamefont {Joo}}]{park2014controlling}%
  \BibitemOpen
  \bibfield  {author} {\bibinfo {author} {\bibfnamefont {J.~H.}\ \bibnamefont
  {Park}}, \bibinfo {author} {\bibfnamefont {V.}~\bibnamefont {Kalra}}, \ and\
  \bibinfo {author} {\bibfnamefont {Y.~L.}\ \bibnamefont {Joo}},\ }\href@noop
  {} {\bibfield  {journal} {\bibinfo  {journal} {The Journal of chemical
  physics}\ }\textbf {\bibinfo {volume} {140}},\ \bibinfo {pages} {124903}
  (\bibinfo {year} {2014})}\BibitemShut {NoStop}%
\bibitem [{\citenamefont {Stephanou}(2015)}]{stephanou2015flow}%
  \BibitemOpen
  \bibfield  {author} {\bibinfo {author} {\bibfnamefont {P.~S.}\ \bibnamefont
  {Stephanou}},\ }\href@noop {} {\bibfield  {journal} {\bibinfo  {journal} {The
  Journal of chemical physics}\ }\textbf {\bibinfo {volume} {142}},\ \bibinfo
  {pages} {064901} (\bibinfo {year} {2015})}\BibitemShut {NoStop}%
\bibitem [{\citenamefont {Yamamoto}\ and\ \citenamefont
  {Kanda}(2012)}]{yamamoto2012computational}%
  \BibitemOpen
  \bibfield  {author} {\bibinfo {author} {\bibfnamefont {T.}~\bibnamefont
  {Yamamoto}}\ and\ \bibinfo {author} {\bibfnamefont {N.}~\bibnamefont
  {Kanda}},\ }\href@noop {} {\bibfield  {journal} {\bibinfo  {journal} {Journal
  of Non-Newtonian Fluid Mechanics}\ }\textbf {\bibinfo {volume} {181}},\
  \bibinfo {pages} {1} (\bibinfo {year} {2012})}\BibitemShut {NoStop}%
\bibitem [{\citenamefont {Hoffmann}\ \emph {et~al.}(1981)\citenamefont
  {Hoffmann}, \citenamefont {Platz}, \citenamefont {Rehage}, \citenamefont
  {Schorr},\ and\ \citenamefont {Ulbricht}}]{hoffmann1981viskoelastische}%
  \BibitemOpen
  \bibfield  {author} {\bibinfo {author} {\bibfnamefont {H.}~\bibnamefont
  {Hoffmann}}, \bibinfo {author} {\bibfnamefont {G.}~\bibnamefont {Platz}},
  \bibinfo {author} {\bibfnamefont {H.}~\bibnamefont {Rehage}}, \bibinfo
  {author} {\bibfnamefont {W.}~\bibnamefont {Schorr}}, \ and\ \bibinfo {author}
  {\bibfnamefont {W.}~\bibnamefont {Ulbricht}},\ }\href@noop {} {\bibfield
  {journal} {\bibinfo  {journal} {Berichte der Bunsengesellschaft f{\"u}r
  physikalische Chemie}\ }\textbf {\bibinfo {volume} {85}},\ \bibinfo {pages}
  {255} (\bibinfo {year} {1981})}\BibitemShut {NoStop}%
\bibitem [{\citenamefont {Rehage}\ and\ \citenamefont
  {Hoffmann}(1982)}]{rehage1982shear}%
  \BibitemOpen
  \bibfield  {author} {\bibinfo {author} {\bibfnamefont {H.}~\bibnamefont
  {Rehage}}\ and\ \bibinfo {author} {\bibfnamefont {H.}~\bibnamefont
  {Hoffmann}},\ }\href@noop {} {\bibfield  {journal} {\bibinfo  {journal}
  {Rheologica Acta}\ }\textbf {\bibinfo {volume} {21}},\ \bibinfo {pages} {561}
  (\bibinfo {year} {1982})}\BibitemShut {NoStop}%
\bibitem [{\citenamefont {F{\"o}rster}\ \emph {et~al.}(2005)\citenamefont
  {F{\"o}rster}, \citenamefont {Konrad},\ and\ \citenamefont
  {Lindner}}]{forster2005shear}%
  \BibitemOpen
  \bibfield  {author} {\bibinfo {author} {\bibfnamefont {S.}~\bibnamefont
  {F{\"o}rster}}, \bibinfo {author} {\bibfnamefont {M.}~\bibnamefont {Konrad}},
  \ and\ \bibinfo {author} {\bibfnamefont {P.}~\bibnamefont {Lindner}},\
  }\href@noop {} {\bibfield  {journal} {\bibinfo  {journal} {Physical review
  letters}\ }\textbf {\bibinfo {volume} {94}},\ \bibinfo {pages} {017803}
  (\bibinfo {year} {2005})}\BibitemShut {NoStop}%
\bibitem [{\citenamefont {Rehage}\ and\ \citenamefont
  {Hoffmann}(1991)}]{rehage1991viscoelastic}%
  \BibitemOpen
  \bibfield  {author} {\bibinfo {author} {\bibfnamefont {H.}~\bibnamefont
  {Rehage}}\ and\ \bibinfo {author} {\bibfnamefont {H.}~\bibnamefont
  {Hoffmann}},\ }\href@noop {} {\bibfield  {journal} {\bibinfo  {journal}
  {Molecular Physics}\ }\textbf {\bibinfo {volume} {74}},\ \bibinfo {pages}
  {933} (\bibinfo {year} {1991})}\BibitemShut {NoStop}%
\bibitem [{\citenamefont {Turner}\ and\ \citenamefont
  {Cates}(1990)}]{turner1990relaxation}%
  \BibitemOpen
  \bibfield  {author} {\bibinfo {author} {\bibfnamefont {M.}~\bibnamefont
  {Turner}}\ and\ \bibinfo {author} {\bibfnamefont {M.}~\bibnamefont {Cates}},\
  }\href@noop {} {\bibfield  {journal} {\bibinfo  {journal} {Journal de
  Physique}\ }\textbf {\bibinfo {volume} {51}},\ \bibinfo {pages} {307}
  (\bibinfo {year} {1990})}\BibitemShut {NoStop}%
\bibitem [{\citenamefont {Zhao}\ \emph {et~al.}(2017)\citenamefont {Zhao},
  \citenamefont {Zhang}, \citenamefont {Zou}, \citenamefont {Dai},
  \citenamefont {Gao}, \citenamefont {Li}, \citenamefont {Lv}, \citenamefont
  {Jiang},\ and\ \citenamefont {Wu}}]{zhao2017can}%
  \BibitemOpen
  \bibfield  {author} {\bibinfo {author} {\bibfnamefont {M.}~\bibnamefont
  {Zhao}}, \bibinfo {author} {\bibfnamefont {Y.}~\bibnamefont {Zhang}},
  \bibinfo {author} {\bibfnamefont {C.}~\bibnamefont {Zou}}, \bibinfo {author}
  {\bibfnamefont {C.}~\bibnamefont {Dai}}, \bibinfo {author} {\bibfnamefont
  {M.}~\bibnamefont {Gao}}, \bibinfo {author} {\bibfnamefont {Y.}~\bibnamefont
  {Li}}, \bibinfo {author} {\bibfnamefont {W.}~\bibnamefont {Lv}}, \bibinfo
  {author} {\bibfnamefont {J.}~\bibnamefont {Jiang}}, \ and\ \bibinfo {author}
  {\bibfnamefont {Y.}~\bibnamefont {Wu}},\ }\href@noop {} {\bibfield  {journal}
  {\bibinfo  {journal} {Materials}\ }\textbf {\bibinfo {volume} {10}},\
  \bibinfo {pages} {1096} (\bibinfo {year} {2017})}\BibitemShut {NoStop}%
\bibitem [{\citenamefont {Shashkina}\ \emph {et~al.}(2005)\citenamefont
  {Shashkina}, \citenamefont {Philippova}, \citenamefont {Zaroslov},
  \citenamefont {Khokhlov}, \citenamefont {Pryakhina},\ and\ \citenamefont
  {Blagodatskikh}}]{shashkina2005rheology}%
  \BibitemOpen
  \bibfield  {author} {\bibinfo {author} {\bibfnamefont {J.~A.}\ \bibnamefont
  {Shashkina}}, \bibinfo {author} {\bibfnamefont {O.~E.}\ \bibnamefont
  {Philippova}}, \bibinfo {author} {\bibfnamefont {Y.~D.}\ \bibnamefont
  {Zaroslov}}, \bibinfo {author} {\bibfnamefont {A.~R.}\ \bibnamefont
  {Khokhlov}}, \bibinfo {author} {\bibfnamefont {T.~A.}\ \bibnamefont
  {Pryakhina}}, \ and\ \bibinfo {author} {\bibfnamefont {I.~V.}\ \bibnamefont
  {Blagodatskikh}},\ }\href@noop {} {\bibfield  {journal} {\bibinfo  {journal}
  {Langmuir}\ }\textbf {\bibinfo {volume} {21}},\ \bibinfo {pages} {1524}
  (\bibinfo {year} {2005})}\BibitemShut {NoStop}%
\bibitem [{\citenamefont {Nettesheim}\ \emph {et~al.}(2008)\citenamefont
  {Nettesheim}, \citenamefont {Liberatore}, \citenamefont {Hodgdon},
  \citenamefont {Wagner}, \citenamefont {Kaler},\ and\ \citenamefont
  {Vethamuthu}}]{nettesheim2008influence}%
  \BibitemOpen
  \bibfield  {author} {\bibinfo {author} {\bibfnamefont {F.}~\bibnamefont
  {Nettesheim}}, \bibinfo {author} {\bibfnamefont {M.~W.}\ \bibnamefont
  {Liberatore}}, \bibinfo {author} {\bibfnamefont {T.~K.}\ \bibnamefont
  {Hodgdon}}, \bibinfo {author} {\bibfnamefont {N.~J.}\ \bibnamefont {Wagner}},
  \bibinfo {author} {\bibfnamefont {E.~W.}\ \bibnamefont {Kaler}}, \ and\
  \bibinfo {author} {\bibfnamefont {M.}~\bibnamefont {Vethamuthu}},\
  }\href@noop {} {\bibfield  {journal} {\bibinfo  {journal} {Langmuir}\
  }\textbf {\bibinfo {volume} {24}},\ \bibinfo {pages} {7718} (\bibinfo {year}
  {2008})}\BibitemShut {NoStop}%
\bibitem [{\citenamefont {Helgeson}\ \emph {et~al.}(2010)\citenamefont
  {Helgeson}, \citenamefont {Hodgdon}, \citenamefont {Kaler}, \citenamefont
  {Wagner}, \citenamefont {Vethamuthu},\ and\ \citenamefont
  {Ananthapadmanabhan}}]{helgeson2010formation}%
  \BibitemOpen
  \bibfield  {author} {\bibinfo {author} {\bibfnamefont {M.~E.}\ \bibnamefont
  {Helgeson}}, \bibinfo {author} {\bibfnamefont {T.~K.}\ \bibnamefont
  {Hodgdon}}, \bibinfo {author} {\bibfnamefont {E.~W.}\ \bibnamefont {Kaler}},
  \bibinfo {author} {\bibfnamefont {N.~J.}\ \bibnamefont {Wagner}}, \bibinfo
  {author} {\bibfnamefont {M.}~\bibnamefont {Vethamuthu}}, \ and\ \bibinfo
  {author} {\bibfnamefont {K.}~\bibnamefont {Ananthapadmanabhan}},\ }\href@noop
  {} {\bibfield  {journal} {\bibinfo  {journal} {Langmuir}\ }\textbf {\bibinfo
  {volume} {26}},\ \bibinfo {pages} {8049} (\bibinfo {year}
  {2010})}\BibitemShut {NoStop}%
\bibitem [{\citenamefont {Fan}\ \emph {et~al.}(2015)\citenamefont {Fan},
  \citenamefont {Li}, \citenamefont {Zhang}, \citenamefont {Fan}, \citenamefont
  {Li},\ and\ \citenamefont {Dong}}]{fan2015nanoparticles}%
  \BibitemOpen
  \bibfield  {author} {\bibinfo {author} {\bibfnamefont {Q.}~\bibnamefont
  {Fan}}, \bibinfo {author} {\bibfnamefont {W.}~\bibnamefont {Li}}, \bibinfo
  {author} {\bibfnamefont {Y.}~\bibnamefont {Zhang}}, \bibinfo {author}
  {\bibfnamefont {W.}~\bibnamefont {Fan}}, \bibinfo {author} {\bibfnamefont
  {X.}~\bibnamefont {Li}}, \ and\ \bibinfo {author} {\bibfnamefont
  {J.}~\bibnamefont {Dong}},\ }\href@noop {} {\bibfield  {journal} {\bibinfo
  {journal} {Colloid and Polymer Science}\ }\textbf {\bibinfo {volume} {293}},\
  \bibinfo {pages} {2507} (\bibinfo {year} {2015})}\BibitemShut {NoStop}%
\bibitem [{\citenamefont {Pletneva}\ \emph {et~al.}(2015)\citenamefont
  {Pletneva}, \citenamefont {Molchanov},\ and\ \citenamefont
  {Philippova}}]{pletneva2015viscoelasticity}%
  \BibitemOpen
  \bibfield  {author} {\bibinfo {author} {\bibfnamefont {V.~A.}\ \bibnamefont
  {Pletneva}}, \bibinfo {author} {\bibfnamefont {V.~S.}\ \bibnamefont
  {Molchanov}}, \ and\ \bibinfo {author} {\bibfnamefont {O.~E.}\ \bibnamefont
  {Philippova}},\ }\href@noop {} {\bibfield  {journal} {\bibinfo  {journal}
  {Langmuir}\ }\textbf {\bibinfo {volume} {31}},\ \bibinfo {pages} {110}
  (\bibinfo {year} {2015})}\BibitemShut {NoStop}%
\bibitem [{\citenamefont {Fei}\ \emph {et~al.}(2017)\citenamefont {Fei},
  \citenamefont {Zhu}, \citenamefont {Xu}, \citenamefont {Li}, \citenamefont
  {Gonzalez},\ and\ \citenamefont {Haghighi}}]{fei2017experimental}%
  \BibitemOpen
  \bibfield  {author} {\bibinfo {author} {\bibfnamefont {Y.}~\bibnamefont
  {Fei}}, \bibinfo {author} {\bibfnamefont {J.}~\bibnamefont {Zhu}}, \bibinfo
  {author} {\bibfnamefont {B.}~\bibnamefont {Xu}}, \bibinfo {author}
  {\bibfnamefont {X.}~\bibnamefont {Li}}, \bibinfo {author} {\bibfnamefont
  {M.}~\bibnamefont {Gonzalez}}, \ and\ \bibinfo {author} {\bibfnamefont
  {M.}~\bibnamefont {Haghighi}},\ }\href@noop {} {\bibfield  {journal}
  {\bibinfo  {journal} {Journal of industrial and engineering chemistry}\
  }\textbf {\bibinfo {volume} {50}},\ \bibinfo {pages} {190} (\bibinfo {year}
  {2017})}\BibitemShut {NoStop}%
\bibitem [{\citenamefont {Mubeena}\ and\ \citenamefont
  {Chatterji}(2015)}]{mubeena2015hierarchical}%
  \BibitemOpen
  \bibfield  {author} {\bibinfo {author} {\bibfnamefont {S.}~\bibnamefont
  {Mubeena}}\ and\ \bibinfo {author} {\bibfnamefont {A.}~\bibnamefont
  {Chatterji}},\ }\href@noop {} {\bibfield  {journal} {\bibinfo  {journal}
  {Physical Review E}\ }\textbf {\bibinfo {volume} {91}},\ \bibinfo {pages}
  {032602} (\bibinfo {year} {2015})}\BibitemShut {NoStop}%
\bibitem [{\citenamefont {{Mubeena}}\ and\ \citenamefont
  {{Chatterji}}(2018)}]{arxive1}%
  \BibitemOpen
  \bibfield  {author} {\bibinfo {author} {\bibfnamefont {S.}~\bibnamefont
  {{Mubeena}}}\ and\ \bibinfo {author} {\bibfnamefont {A.}~\bibnamefont
  {{Chatterji}}},\ }\href@noop {} {\bibfield  {journal} {\bibinfo  {journal}
  {ArXiv e-prints}\ } (\bibinfo {year} {2018})},\ \Eprint
  {http://arxiv.org/abs/1801.06933} {arXiv:1801.06933 [cond-mat.soft]}
  \BibitemShut {NoStop}%
\bibitem [{\citenamefont {{Mubeena}}(2018{\natexlab{a}})}]{arxive2}%
  \BibitemOpen
  \bibfield  {author} {\bibinfo {author} {\bibfnamefont {S.}~\bibnamefont
  {{Mubeena}}},\ }\href@noop {} {\bibfield  {journal} {\bibinfo  {journal}
  {ArXiv e-prints}\ } (\bibinfo {year} {2018}{\natexlab{a}})},\ \Eprint
  {http://arxiv.org/abs/1806.02509} {arXiv:1806.02509 [cond-mat.soft]}
  \BibitemShut {NoStop}%
\bibitem [{\citenamefont {{Mubeena}}(2018{\natexlab{b}})}]{arxive3}%
  \BibitemOpen
  \bibfield  {author} {\bibinfo {author} {\bibfnamefont {S.}~\bibnamefont
  {{Mubeena}}},\ }\href@noop {} {\bibfield  {journal} {\bibinfo  {journal}
  {ArXiv e-prints}\ } (\bibinfo {year} {2018}{\natexlab{b}})},\ \Eprint
  {http://arxiv.org/abs/1806.02504} {arXiv:1806.02504 [cond-mat.soft]}
  \BibitemShut {NoStop}%
\bibitem [{\citenamefont {{Mubeena}}(2018{\natexlab{c}})}]{arxive4}%
  \BibitemOpen
  \bibfield  {author} {\bibinfo {author} {\bibfnamefont {S.}~\bibnamefont
  {{Mubeena}}},\ }\href@noop {} {\bibfield  {journal} {\bibinfo  {journal}
  {ArXiv e-prints}\ } (\bibinfo {year} {2018}{\natexlab{c}})},\ \Eprint
  {http://arxiv.org/abs/1809.05294} {arXiv:1809.05294 [cond-mat.soft]}
  \BibitemShut {NoStop}%
\end{thebibliography}%

\end{document}